\documentclass[superscriptaddress,prd,preprintnumbers,nofootinbib,aps,9pt]{revtex4}
\usepackage{graphics, epstopdf}

\usepackage{subeqnarray}
\usepackage{epsfig,amsmath,amssymb}
\usepackage{mathrsfs}
\usepackage[usenames,dvipsnames]{color}
\usepackage[pagebackref=true, colorlinks=true]{hyperref}
\definecolor{redish}{rgb}{0.7,0.2,0.0} % color defined in (r=red,g=green,b=blue) model
\definecolor{bluish}{rgb}{0.2,0.5,0.8}
\hypersetup{linkcolor=redish, % color of internal links
 citecolor=blue, % color of links to bibliography
 filecolor=magenta, % color of file links
 urlcolor=bluish} 

\usepackage{amsmath} % Useful for equation alignment
\usepackage{verbatim} % Allows to have comments over multiple lines
\usepackage{mathrsfs}
\usepackage{amsfonts}
\usepackage[english]{babel}
\usepackage[utf8]{inputenc}
\usepackage[T1]{fontenc}
\usepackage{hyperref}
\newcommand{\tc}{\eta_{k}^c}
\newcommand{\beq}{\begin{equation}}
\newcommand{\eeq}{\end{equation}}
\newcommand{\nk}{\vec{k}}
\newcommand{\nq}{\vec{q}}
\newcommand{\dphi}{\delta \phi}
\newcommand{\x}{\vec{x}}

\newcommand{\bra}{\langle}
\newcommand{\ket}{\rangle}
\newcommand{\bea}{\begin{eqnarray}}
\newcommand{\eea}{\end{eqnarray}}
\newcommand{\mH}{\mathcal{H}}
\newcommand{\mP}{\mathcal{P}}

\newcommand{\nn}{\nonumber \\}
 \newcommand{\Ny}{N^{(y)}}
\newcommand{\BB}{\textrm{BB}}
\newcommand{\CSL}{\textrm{CSL}}
\newcommand{\Nt}{\textrm{Newt}}
 
 \newcommand{\avg}[1]{\mathbb{E} \{ {#1} \}}
 \newcommand{\expec}[1]{\langle {#1} \rangle}
 \newcommand{\vp}{\vec{p}\:}
 \newcommand{\vkappa}{\vec{\kappa}\:}
 \newcommand{\vq}{\vec{q}\:}
 \newcommand{\vk}{\vec{k}\:}
 \newcommand{\RI}{\textrm{R,I}} 
\newcommand{\yk}{y_{\vec{q}}}
\newcommand{\pk}{\pi_{\vec{q}}} 
 
 \newcommand{\1}{{(1)}}
 \newcommand{\2}{{(2)}}
 \def \deu{{\delta}^{(1)}}
 \def \ded{{\delta}^{(2)}}

 \def\be{\begin{equation}}
 \def\ee{\end{equation}}
 \def\bea{\begin{eqnarray}}
 \def\eea{\end{eqnarray}}
 \def\Aa{\frac{\dot a}{a}}
 \def\Ab{\Big(\frac{\dot a}{a}\Big)^2}
 \def\Ac{\frac{\ddot a}{a}}

 \def\La{\partial_i \,\partial^i}
 \def\LA{\partial_k \,\partial^k}
 \def\deu{{\delta}^{(1)}}
 \def\ded{{\delta}^{(2)}}

\newcommand{\ba}{\begin{eqnarray}}
\newcommand{\ea}{\end{eqnarray}}
\newcommand{\bii}{\begin{itemize}}
\newcommand{\eii}{\end{itemize}}
\newcommand{\f}{\frac}
\def \p{\partial}
\def \d{\delta}

\def \a{\alpha}
\def \s{\sigma}

\def \r{\rho}

\def \n{\eta}
\def \({\left(}
\def \){\right)}
\def \[{\left[}
\def \]{\right]}
\def \bg{\bar{g}}

\def \v{\vec}

\begin{document}

 \title{On the expectation of primordial gravity waves generated during inflation} 
 
 \author{Gabriel León}
\email{gleon@fcaglp.unlp.edu.ar}
\affiliation{Grupo de Astrof\'isica, Relatividad y 
Cosmolog\'ia, Facultad de Ciencias 
Astron\'omicas y Geof\'isicas, Universidad Nacional de La Plata, Paseo del Bosque S/N 
(1900) La Plata, Argentina.\\
CONICET, Godoy Cruz 2290, 1425 Ciudad Aut\'onoma de Buenos Aires, Argentina.}
\author{Abhishek Majhi}
\email{abhishek.majhi@gmail.com}
\affiliation{Instituto de Ciencias Nucleares, Universidad Nacional Aut\'onoma de M\'exico, Mexico City, 04510, Mexico.}
\author{Elias Okon}
\email{eokon@filosoficas.unam.mx}
\affiliation{Instituto de Investigaciones Filos\'oficas, Universidad Nacional Aut\'onoma 
de M\'exico, Mexico City, 04510, Mexico.}
\author{Daniel Sudarsky}
\email{sudarsky@nucleares.unam.mx}
\affiliation{Instituto de Ciencias Nucleares, Universidad Nacional Aut\'onoma de M\'exico, Mexico City, 04510, Mexico.}

\begin{abstract}
The inflationary paradigm is extremely successful regarding predictions of temperature anisotropies in the CMB. However, inflation also makes predictions for a CMB B-mode polarization, which has \emph{not} been detected. Moreover, the standard inflationary paradigm is unable to accommodate the evolution from the initial state, which is assumed to be symmetric, into a non-symmetric aftermath. In \cite{USshort}, we show that the incorporation of an element capable of explaining such a transition drastically changes the prediction for the shape and size of the B-mode spectrum. In particular, employing a realistic objective collapse model in a well-defined semi-classical context, we find that, while predictions of temperature anisotropies are nor altered (with respect to standard predictions), the B-mode spectrum gets strongly suppressed---in accordance with observations. Here we present an in-depth discussion of that analysis, together with the details of the calculation. %We also argue that comparable results are to be expected even within a fully quantum treatment of gravity.
\end{abstract}

\maketitle
%%%%%%%%%%%%%%%%%%%%%%%%%%%%%%%%%%%%%%%%%%%%%%%%%%%%%%%%%%%%%
%%%%%%%%%%%%%%%%%%%%%%%%%%%%%%%%%%%%%%%%%%%%%%%%%%%%%%%%%%%%%
\section{Introduction}
\label{intro}
%%%%%%%%%%%%%%%%%%%%%%%%%%%%%%%%%%%%%%%%%%%%%%%%%%%%%%%%%%%%%
%%%%%%%%%%%%%%%%%%%%%%%%%%%%%%%%%%%%%%%%%%%%%%%%%%%%%%%%%%%%%

The measurement problem has bothered physicists since the birth of quantum theory. In short, the problem consists of the fact that the standard formalism crucially depends on notions such as \emph{measurement} or \emph{observer} (to decide when to use the unitary evolution and when the collapse postulate), but such notions are never formally defined within the theory (i.e., the formalism fails to offer a detailed and unambiguous prescription identifying the interactions and objects that should be taken as playing such roles) \cite{Bell2,bell,Maudlin}. One of the few viable approaches to tackle the issue involves the incorporation of spontaneous wave function collapses \cite{grw,per,gpr};\footnote{See \cite{Maudlin} for a helpful clasiffication of the set of viable approaches.} the idea is to come up with a unified dynamics that encompasses both the unitary evolution and the collapse mechanism. In the Continuous Spontaneous Localisation (CSL) approach \cite{per,bassi}, this is done mathematically by modifying the unitary Schrödinger evolution with the introduction of specific non-linear, stochastic terms designed to drive any initial wave function into one of the eigenstate of a, so-called, collapse operator. As a result, within this scheme, one does not depend on an {\it ad hoc} identification of observers or measurements in order to apply the formalism and explicit predictions can be obtained regarding situations, such as cosmology, where no observers or measuring apparatuses can be identified (thus removing the well-known conceptual obstacles for the application of quantum theory to such a field).\footnote{The problems one faces when attempting to apply standard quantum mechanics to cosmology motivated the developement of the so-called, Consistent Histories approach \cite{CH1,CH2,CH3} (Consistent Histories was independently introduced in \cite{Gri:84,Gri:86,Gri:87,Omn:87,Omn:88}). Unfortunately, under closer inspection such a formalism was shown to contain problems that render it unviable \cite{CHus1,CHus2,CHus3}.}

Since the construction of quantum field theory is based on standard quantum mechanics, a modification of the latter, such as the one proposed by CSL, clearly affects the former. Given that both standard quantum mechanics and quantum field theory are very well-tested, one may wonder if such a modification can be done without disturbing their empirical success. Maybe surprisingly, the answer is in the affirmative \cite{Bas:13}. One may also wonder if a modification of this kind is really necessary. The answer, again, is yes, especially if one takes quantum theory to be fundamental and thinks that the notion of observer should arise from within the theory and not as something external. Of course, such a point of view is essential for an application of quantum field theory to a cosmological setting, especially while studying the early universe: in such a scenario, one clearly cannot rely on the notion of an external observer. Therefore, in order to apply quantum field theory to an inflationary era, an observer-independent quantum dynamics, such as CSL, seems obligatory \cite{pss,gs1,gs2,ppd,ts,gs3}.

In order to focus more sharply on the motivation for a modified framework, let us consider the standard approach to inflationary cosmology. In such a treatment, the background space-time is taken to be a Friedman-Robertson-Walker (FRW) universe, whose expansion is driven by a scalar field called the \emph{inflaton} \cite{wein}. The initial state of this field is assumed to be the homogeneous and isotropic Bunch-Davies vacuum and the \emph{quantum fluctuations} of this state are regarded as seeds for the anisotropic and inhomogeneous cosmic structures of the present universe. However, this passage from quantum fluctuations to actual structure is questionable, or at least incomplete. The problem is that such fluctuations or uncertainties cannot be taken to represent \emph{physical} fluctuations, they are only a measure of the \emph{width} of the quantum state in question.

To see this, consider the ground state of a 1D simple harmonic oscillator, which of course has uncertainty in position. Note however that such an uncertainty does \emph{not} imply that the ground state is not symmetric under a reflection $x\to-x$; instead, the uncertainty is a measure of the spread of the results of several position measurements, performed on an ensemble of identically prepared systems. Therefore, in order to break the reflection symmetry of a single harmonic oscillator, an actual \emph{measurement} of position has to be performed. In other words, the quantum fluctuations or uncertainties do not, by themselves, indicate that some aspect of the physical system is undergoing random motion, and as far as a quantum state of the system is taken to describe it completely, the symmetries of the quantum state must be taken as also characterizing the system to which such a state is associated. Similarly, the fluctuations or uncertainties in the Bunch-Davies vacuum do not, in any way, constitute a departure from homogeneity or isotropy. Without an actual, physical change, beyond that imposed by the unitary dynamics (which clearly does not break such symmetries), no deviation from the initially symmetric state can occur. And since, as we discussed above, no measurements can happen in this setting, clearly there is something missing in the inflationary account of the emergence of seeds of cosmic structure. This issue can be taken care of by employing an objective collapse model, such as CSL. In that case, the passage from a homogeneous and isotropic state to an inhomogeneous and anisotropic outcome occurs via the physical process of wave function collapse, without the need of an intervention by any observer. 

There are, then, enough physical and conceptual motivations, both from the perspective of quantum theory and that of cosmology, to consider a modified quantum theory that introduces objective, spontaneous collapses of the wave function. In this paper we explore the consequences of adopting such a point of view for the prediction of primordial gravity waves generated during inflation. We do so because, while the inflationary paradigm is extremely successful regarding predictions of temperature anisotropies in the CMB, such paradigm also makes predictions for an observable CMB B-mode polarization. The problem is that, to date, such polarization has \emph{not} been detected,\footnote{As is well-known, recent hopes of such detection got nullified by dust polarization \cite{dust1,dust2,dust3,dust4,dust5}.} and that fact has been used to rule-out some of the simplest models of inflation \cite{jmartin2013,jmartin2014,planck2015inflation}. In \cite{glsus,finl} it was shown that the incorporation of a rudimentary objective collapse mechanism leads to a highly suppressed amplitude of the B-mode spectrum. Then, in \cite{USshort} we put such a result on an even stronger ground by obtaining analogous results with the adaptation of a \emph{realistic} objective collapse model to the situation at hand. 

In this manuscript we present an in-depth discussion of that analysis, together with the details of the calculation. For these purposes the paper is organized as follows. In section \ref{semi-classical}, we present the general conceptual framework that underlies our approach, indicating how the objective collapse of the wave function can be incorporated into our general understanding of the gravity-quantum interface. Next, in section \ref{theorypert} we present a technical summary of results within cosmological perturbation theory during inflation that will be relevant for our work. Afterwards, in section \ref{secPStensor} we describe how the self-induced collapse of the wave function generates the primordial gravitational waves and show how such a proposal leads to a strong suppression in the estimate for the amplitude of the spectrum corresponding to the primordial tensor modes. In order to illustrate the generality of our results, we do so both using a realistic CSL collapse mechanism and a simplistic toy model of collapse. In section \ref{predictions} we discuss in detail how our model affects the predictions for the usual observables related to primordial gravity waves. Finally, in section \ref{concl} we provide a brief summary of the results and our conclusions (we also include an appendix where the details of the calculations can be found).

%%%%%%%%%%%%%%%%%%%%%%%%%%%%%%%%%%%%%%%%%%%%%%%%%%%%%%%%%%%%%
%%%%%%%%%%%%%%%%%%%%%%%%%%%%%%%%%%%%%%%%%%%%%%%%%%%%%%%%%%%%%
\section{Objective Collapse in a Semi-classical Setting}
\label{semi-classical}
%%%%%%%%%%%%%%%%%%%%%%%%%%%%%%%%%%%%%%%%%%%%%%%%%%%%%%%%%%%%%
%%%%%%%%%%%%%%%%%%%%%%%%%%%%%%%%%%%%%%%%%%%%%%%%%%%%%%%%%%%%%

The inflationary account of the emergence of cosmic structure, via quantum fluctuations, forces us to face head-on the quantum-gravity interface. In fact, as emphasized in \cite{jmartin,jmartin2015}, such a situation is the only one that, at this point, brings together quantum theory, general relativity and observations. The situation is rather delicate from both the conceptual and technical points of view. On the technical side, we must recognize that, despite heroic efforts and advances made in the various programs searching for a quantum theory of gravity, we currently do not have a mature, fully workable theory deserving that name. Such a state of affairs is evident by our inability to answer questions such as: what would be the space-time associated with a large massive body, for instance a 1 ton ball of iron, in a quantum superposition of two widely separate spatial locations? Ideally, one would be able to produce some kind of state in a suitable Hilbert space characterizing the quantum superposition of space-time metrics. Unfortunately, as far as we know, none of the quantum gravity programs, as of today, can provide a satisfactory answer to that situation. On the other hand, at the conceptual level, one encounters thorny issues, such as the well-known problem of time afflicting canonical quantum theories of gravity and the issue of recovering from the fundamental theory classical space-time notions through suitable approximations. Moreover, one probably has to take a stance regarding the often overlooked conceptual problems within the foundations of quantum theory mentioned above.
 
Most works addressing quantum aspects of inflationary cosmology simply set aside all these conceptual questions, which are considered, at best, irrelevant subtleties or, at worst, annoying distractions. Such an attitude is understandable, given the very small amplitude of the quantum fluctuations (which we might think of as characterized by the $10^{-5}$ amplitude in the CMB temperature fluctuations). This seems to imply that, irrespective of the precise way in which the conceptual difficulties are handled, the quantum aspects of space-time will only induce very small departures from the causal structure of the classical background metric. Further impetus for ignoring conceptual issues comes from the undeniable phenomenological success of the whole enterprise regarding scalar perturbations. Nevertheless, as already discussed in full detail in, e.g., \cite{Shortcomings}, the important question of accounting for the transition from a fully homogeneous state to one containing actual inhomogeneities has not found a satisfactory answer withing the standard treatment.

In order to deal with the aforementioned issues, we follow a program based on \emph{semi-classical gravity}. This is a framework in which matter fields are fully quantum, but space-time is fully classic. This might sound natural at first, but raises serious questions after some thought. It is worth noting, though, that this is probably the best we can hope to do with reasonable rigor, given the fact that, as previously mentioned, we do not have a complete theory of quantum gravity. Moreover, we are of course not the first to consider such an approach, which has a rather long history and substantial literature behind it \cite{DeWitt1975,Eppley1977,Birrell,kiefer1993,Wald94}. There is, however, a very influential work, \cite{Page}, which is often taken to rule-out semi-classical gravity all together. The argument, based on an actual experiment attempting to create a superposition of two space-time metrics, holds that semi-classical gravity, without a collapse of the quantum state, leads to predictions that are in conflict with observations, but that the introduction of a collapse leads to a violation of the semi-classical equation $ G_{ab } = 8\pi G \bra \hat T_{ab} \ket$ (because the LHS has vanishing divergence but the RHS, as a result of the collapses, would have non-zero divergence). Either way, the theory is in trouble. 
 
We acknowledge that the argument described above represents a serious obstacle for the consideration of semi-classical gravity as a \emph{fundamental} description of the situation at hand. Nonetheless, we do not see it as an impediment for taking semi-classical gravity, supplemented by a modified quantum dynamics involving spontaneous collapse of the wave function, as a viable \emph{effective} description, with restricted but rather wide applicability. The idea is to regard semi-classical gravity in analogy with hydrodynamics. We know that hydrodynamics, as described with the Navier-Stokes equations (NSE), constitutes a mathematically sound theory, which provides a rather robust description of fluids under a rather wide set of circumstances. Yet, we know that the NSE do not provide a fundamental description of actual fluids. At a deeper level, there is a molecular or atomic characterization of the elements that make up the fluid and the forces between them. As a result, basic concepts of fluid dynamics, such as velocity, pressure, density or viscosity (not to mention other complex properties such as vorticity and laminar flow) are not present at the level of the deeper description; those properties are clearly emergent notions that are adequate for the effective description of the system under limited circumstances. At the same time, due to the fact that the NSE are recognized as non-fundamental, we are not likely to react with surprise or disbelief if there are conditions where the NSE fail or do not even make sense.

Consider, for instance, a wave in the ocean that, when propagating smoothly, is treated according to the NSE. When such a wave reaches the beach and breaks down, the description of the situation requires a treatment that goes well beyond what the NSE can provide. In fact, it is clear that, under certain conditions, the very notion of a fixed fluid volume, with a definite 3-velocity and mean density, simply ceases to make sense. We are adopting a similar point of view regarding gravity. That is, the characterization of space-time via a smooth pseudo-Riemannian metric obeying Einstein's semi-classical equations is taken as analogous to the description of fluids via the NSE, i.e. one views the notion of space-time as emergent from deeper, probably not even geometrical (in the Riemannian sense), quantum gravity degrees of freedom. The semi-classical equation $ G_{ab } = 8\pi G \bra \hat T_{ab} \ket$ is viewed, thus, as just a good approximate characterization, valid in certain circumstances, and therefore, departures from it, both large and small, should not be taken as surprising. We believe furthermore that the collapses we are introducing correspond, in a sense, to relatively small violations of Einstein's equations. Specifically, just as the NSE can be taken as valid just before and after the breaking of the wave), so can the collapse process be incorporated into the semicalssical treatment, when a judicious gluing process is used to maintain the approximate validity. The formalism of that gluing process is introduced in \cite{ts}, and in the following we will describe briefly the main idea.

We propose, then, to take semi-classical gravity, as described by Einstein semi-classical equations $ G_{ab } = 8\pi G \bra \hat T_{ab} \ket$, together with a quantum dynamics supplemented with an objective collapse mechanics, as an effective description of the interaction between gravity and matter fields, suitable for a large set situations---including inflationary cosmology. This position not only deals with the objections against semi`classical gravity mentioned before, but allows us to provide a clear resolution to the issue of the transition from the homogeneous initial state to one containing the actual seeds of cosmic structure (see \cite{pss,gs1,gs2,ppd,ts}). Also, our proposal leads to the derivation of a spectrum of primordial fluctuations compatible with the observations in the CMB, \cite{gs3,pia,claudia,micol,micol2} (applications of this approach to situations involving black holes have been shown to provide attractive accounts of the so-called \textit{information loss paradox} \cite{BH1,BH2,Elias,BH4,BH5,BH6}, and also different conclusions from the standard approach to the \textit{eternal inflation} scenario \cite{eternalcsl}). 

A detailed formalism realizing these general ideas was introduced in \cite{ts} under the name of the Semi-classical Self-Consistent formalism. The staring point is the notion of a Semi-classical Self-Consistent Configuration (SSC), which is defined as follows. A set $\lbrace g_{ab}(x),\hat{\varphi}(x), \hat{\pi}(x), {\cal H}, \vert \xi \rangle \in {\cal H}\rbrace $ is a SSC if and only if $\hat{\varphi}(x)$, $\hat{\pi}(x)$ and $ {\cal H}$ correspond a to quantum field theory for the field $\varphi(x)$, constructed over a space-time with metric $g_{ab}(x)$, and the state $\vert\xi\rangle$ in $ {\cal H}$ is such that:
\begin{equation}\label{scc}
G_{ab}[g(x)]=8\pi G\langle\xi\vert \hat{T}_{ab}[g(x),\hat{\varphi}(x)]\vert\xi
\rangle .
\end{equation}
This is a natural general-relativistic version of the Schr\"odinger-Newton equation \cite{Diosi1984}, in which one considers the Schr\"odinger equation for a particle, subject to a gravitational field generated by considering the wave function of such a particle as a mass distribution. That is,
\begin{equation}
i\frac{\partial \psi }{\partial t}= -\frac{1}{2M} \nabla^2\psi + M\Phi_N\psi
\end{equation}
 and
\begin{equation}
 \nabla^2\Phi_N = 4\pi G M |\psi|^2.
\end{equation}

In order to incorporate a collapse mechanism to the SSC picture, we consider first the simplest case corresponding to a single, instantaneous jump in the state of the quantum field. Following the GRW prescription, \cite{grw}, we supplement the standard smooth unitary evolution with an objective, spontaneous jump of the quantum state,
\begin{equation}
 \vert \xi \rangle \to \vert \xi^{ \textrm{post-collape}} \rangle.
 \end{equation}
To describe this modified dynamics, we take the Hamiltonian part of the evolution and absorb it in the quantum field operators (as in the standard Heisenberg picture). The reminder of the evolution law, provided by the spontaneous collapses, is treated as an interaction (following the interaction picture approach).

In more detail, in order to combine the SSC formalism and the spontaneous collapses, we start with an initial SSC (which we call SSC1) and we demand the theory to provide, in a stochastic manner, i) a space-like hypersurface of the space-time of SSC1, $\Sigma_{\textrm{Collapse}}$, on which the collapse of the quantum state takes place, and ii) the post-collapse quantum state. Then, with such information, we construct a new SSC (which we call SSC2), which describes the situation after the collapse. Finally, we specify how SSC1 and SSC2 are to be joined in order to generate a ``global space-time.'' Note however that the Hilbert spaces of SSC1 and SSC2 do not coincide. Therefore, in order to construct the post-collapse state of SSC2, we need to first collapse $\vert\xi ^{(1)}\rangle$, the state of SSC1, into a so-called target state also in the Hilbert space of SSC1, $\vert\chi^{t}
\rangle$, and then we use such a target state to construct the actual post-collapse state of SSC2, $\vert\xi ^{(2)}\rangle$. The specific proposal for this construction is the following. First, SSC2 is required to have an hypersurface isometric to $\Sigma_{\textrm{Collapse}}$. Such an hypersurface is where the two space-times are to be joined. Then, to construct the post-collapse state of SSC2 out of the target state, we demand that, on $\Sigma_{\textrm{Collapse}}$,
\begin{equation}\label{scc-joining}
\langle\chi^{t}\vert \hat{T}^{(1)}_{ab}[g(x),\hat{\varphi}(x)]\vert\chi^{t}
\rangle =\langle\xi^{(2)}\vert \hat{T}^{(2)}_{ab}[g(x),\hat{\varphi}(x)]\vert\xi ^{(2)}\rangle ,
\end{equation}
where $\hat{T}^{(1)} $ and $\hat{T}^{(2)} $ are the renormalized energy-momentum tensors of SSC1 and SSC2.

The next step is to construct the full space-time of SSC2, from which the QFT over it can be developed. For this, we note that the space-time metric of SSC1 allows us to construct an induced spatial metric $h_{ab}^{(1)}$ on $\Sigma_{\textrm{Collapse}}$, with unit normal $ n^{a (1)}$ and extrinsic curvature ${K^{ab}}^{(1)}$. Next, out of this data, we need to obtain suitable initial conditions for the space-time metric of SSC2. That is, we need to find a $h_{ab}^{(2)}$ and ${K^{ab}}^{(2)}$ satisfying the Hamiltonian and momentum constraints of SSC2. To do this, we have set $h_{ab}^{(2)} = h_{ab}^{(1)}$ on $ \Sigma_{\textrm{Collapse}}$ and looked for a suitable $K_{ab}^{(2)} $ that satisfied the constraints. After the determination of the initial data for the SSC2 metric, we construct the whole metric by solving the evolution equations of general relativity, together with the conservation equation for $\langle\xi_{(2)}^{phys}\vert \hat{T}_{bc}[g(x),\hat{\varphi}(x)]\vert \xi_{(2)}^{phys} \rangle^{(2)}$. Note that, as previously indicated, such conservation equation will hold after the collapse. An explicit example showing the completion of this process in the inflationary cosmological context was presented in \cite{ts}. There one can see that, in general, the tasks involved in these constructions are rather non-trivial.
 
The former scheme allows for the construction of a ``space-time'' composed of two 4-dimensional regions, each part of a SSC constructions, joined along a collapse hypersurface. By construction, Einstein's semi-classical equations hold in the interior of the space-time regions corresponding to each SSC. However, in the same way that the NSE are not satisfied during the break of an ocean wave, they do not hold on the collapse hypersurface. This formalism might seem rather different than frameworks previously considered. However, the fact is that, in a hidden manner, it underlies other approaches. One such example is the stochastic gravity formalism of \cite{Verdaguer2008,Calzetta1994}. In order to see the connection, consider the following collapse, $\vert\psi (t) \rangle = \theta(t_0-t) \vert 0 \rangle + \theta(t-t_0 ) \vert\xi \rangle$, and analyze its gravitational effects. In this case, the Einstein semi-classical equations can be written as
\begin{equation}\label{Hu}
G_{ab} = 8\pi G \langle 0\vert \hat{T}_{ab}\vert0\rangle,
 + 8\pi G \xi_{ab},
\end{equation}
where 
$\xi_{ab} \equiv \theta(t-t_0 )( \langle\xi\vert \hat{T}_{ab}\vert\xi\rangle-\langle 0\vert \hat{T}_{ab}\vert 0\rangle)$ might be seen as corresponding to an individual stochastic step. As a consequence, stochastic gravity might correspond to a continuous version of a collapse model, such as CSL. In fact, Eq. (\ref{Hu}) has precisely the
 form of the modified semi-classical gravity equation considered in \cite{Verdaguer2008}, where the term $ \xi_{ab}$ is taken to represent a ``stochastic realization'' of the quantum uncertainty of the energy-momentum tensor, as characterized by 
 $\langle\xi\vert \hat{T}_{ab} ( x) \hat{T}_{cd}(y)\vert\xi\rangle- \langle\xi\vert \hat{T}_{ab} ( x)| \xi \rangle \langle \xi | \hat{T}_{cd}(y)\vert\xi\rangle$. We note that, in such context, just as in our own scheme, the fundamental equation cannot be taken to be valid at the ``time of the stochastic jump,'' precisely because of the conflict between the Bianchi identities and the fact that, generically, $ \nabla^a \xi_{ab } \not =0 $ (even if at the level of the average over the ensemble of possible realizations of the stochastic variables such equation holds). However, the equation might well be valid both before and after the stochastic jump.
 
As we have seen, the general application of this formalism is a highly non-trivial task. However, as shown in \cite{ts}, the inflationary cosmology case we will be considering in what follows can be well approximated by maintaining the characterization of the QFT in a single Hilbert space. This simplifies the treatment substantially and allows for a direct extension to theories involving continuous collapse processes, which can be regarded as a succession of infinitesimal steps of the kind described above. This justifies the use of CSL theory in this context and validates its success described in \cite{ppd} in recovering the (almost) scale free spectrum of primordial scalar perturbations that matches the observations of the CMB.

It is important to mention that, as a consequence of the semi-classical gravity framework we follow, our treatment of the scalar and tensor perturbations (of the metric and matter fields during inflation) will be different from the traditional one. In the standarad teratments one encounters a scalar perturbation mode which is made of two parts, one corresponding to a the metric perturbation and one to the perturbation of the inflaton field. The tensor perturbation, on the other hand, corresponds only to aspects of the metric perturbation (and contains no scalar filed perturbation). In the standard approach, one subjects the scalar and tensor modes, which involve inflaton filed and metric perturbations, to a quantum treatment, leading to an analogous treatment for both the scalar and tensor degrees of freedom. By contrast, in our approach, matter fields (both background and perturbations) are in principle treated quantum mechanically (in fact, in \cite{ts} it is shown that one can replace the quantum treatment of the zero mode by the quantum expectation of the field's zero mode), while the metric (both the background and the perturbations) is always considered in a classical manner. This means that the part of the scalar mode corresponding to the inflaton field perturbation is subject to a quantum treatment while, for the tensor modes, there is no part to be treated quantum mechanically. The result is that the collapse of the quantum state of the inflaton field is now the source of both scalar and tensor metric perturbations. In the first case, the source term (appearing in the energy-momentum tensor) is linear both on the zero mode of the inflaton and in the scalar field perturbation. In the second case, the source terms (in the energy-momentum tensor) are quadratic in the perturbation of the scalar field (with additional terms that are quadratic in the scalar perturbations of the metric itself).

%%%%%%%%%%%%%%%%%%%%%%%%%%%%%%%%%%%%%%%%%%%%%%%%%%%%%%%%%%%%%
%%%%%%%%%%%%%%%%%%%%%%%%%%%%%%%%%%%%%%%%%%%%%%%%%%%%%%%%%%%%%
\section{Scalar and tensor perturbations during inflation}\label{theorypert}
%%%%%%%%%%%%%%%%%%%%%%%%%%%%%%%%%%%%%%%%%%%%%%%%%%%%%%%%%%%%%
%%%%%%%%%%%%%%%%%%%%%%%%%%%%%%%%%%%%%%%%%%%%%%%%%%%%%%%%%%%%%

We now provide a brief summary of cosmological perturbation theory, focusing on results that will be of interest for our approach. In what follows, we shall use Greek letters, $\mu,\nu$, etc. to denote space-time indices (they can take values $0,1,2,3$) and Latin letters $i,j,k$, etc. to denote spatial indices (they can take values $1,2,3$). Also, all quantities corresponding to a fixed background space-time will carry a $\bar{~}$. We also take $c=\hbar=1$; hence, the dimensions of mass M, length L and time T are related as $M=1/L=1/T$. This implies that momentum has units of $1/L$ and that the gravitational constant $G$ those of $L^2$.

The simplest model of inflation is described by a single scalar field $\phi$, the inflaton, with standard kinetic energy term, minimal coupling to gravity and potential $V$; the corresponding action is 
\begin{equation}\label{action}
 S = \int d^4 x \sqrt{-g} \left[ \frac{1}{16 \pi G} R[g] - \frac{g^{ab}}{2} \nabla_a \phi \nabla_b \phi - V(\phi) \right].
\end{equation}
The equation of motion for the the inflaton is
% found by setting $\d S/\d \phi=0$ and by imposing the appropriate boundary conditions at asymptotic infinity. This yields
\ba
g^{\mu\nu}\nabla_{\mu}\nabla_{\nu}\phi-\p V/\p \phi=0,
\ea
where $\nabla_{\mu}$ is the covariant derivative compatible with $g_{\mu\nu}$. In order to perform a perturbative treatment, one splits both the metric and the scalar field into backgrounds and fluctuations. The background space-time is characterized by a spatially flat FRW solution and that of the inflaton by its homogeneous part $\phi_0 (\eta)$. We write the background metric as $\bg_{\mu\nu} = a(\eta) \eta_{\mu \nu}$, where $\eta_{\mu \nu}$ is the Minkowski metric in standard coordinates and $a(\eta)$ is the scale factor (with $\eta$ a conformal time). For the background space-time and field, and
with the slow-roll regime (described by $\dot \phi_0 \simeq -(a^3/3 \dot a) \partial_\phi V$) the theory leads to Friedman equations that read $\mH^2 \simeq (8 \pi G /3 ) a^2 V $, where $\mH \equiv \dot a/a$ (the $\dot{~}$ represents derivative with respect to $\eta$). A useful quantity characterizing the ``slowness'' of slow-roll inflationary regime is the Hubble slow-roll parameter $\epsilon \equiv 1- \dot{\mH}/\mH^2$, which during inflation is taken as approximately a constant and $\epsilon \ll 1$. 

Here we need to point out that the above characterization is often considered as referring to the classical aspect of inflationary cosmology, and distinguished from the quantum aspect, reserved to deal with the perturbations. In our approach, based on a semi-classical treatment, the separation between quantum and classical aspects is in principle placed at a different point: The space-time is always described in classical terms while the inflaton field is always described in the language of quantum field theory. Thus, the above characterization must be taken as referring to a situation where the state of the inflaton field is such that the modes involving any space-time dependence are not excited; i.e. are characterized by the vacuum state, (taken as usual as the Bunch Davies vacuum or similar state) while the zero mode (the mode that is homogeneous) is in a highly exited state taken to be something like a coherent state. The details of this construction can be found for instance in \cite{ts}.

The other aspect that needs clarification in our approach is that we are trying to construct an account for the emergence of primordial perturbations that clearly identifies the mechanism and time sequence of the various stages, and in particular describes the emergence as a process occurring in time.\footnote {After all the word ``emergence'' is taken to indicate that something that was not present at an early time, is present at a latter time. } Thus in our approach we seek an account where the early stages of inflation are completely homogeneous and isotropic, with the space-time metric taken to be characterized exactly by the spatially flat FRW solution sourced by an inflation field which is itself in the completely homogeneous and isotropic state characterized in the above paragraph (and described in detail in \cite{ts}). The source of all inhomogeneities and anisotropies is taken to reside in the spontaneous collapse of the quantum state of the field. The inhomogeneities and anisotropies are then transmitted to the space-time geometry as a result of its coupling via Einstein's semi-classical equation to the expectation value of the energy momentum tensor of the quantum field, which develops inhomogeneities as a consequence of the appearance of such features in the quantum state.
 
The above point of view then clearly calls for the study of how the changes in the quantum state of the inflaton field lead to changes in the space-time metric. The changes usually described in terms of the ``Newtonian potential'' are the result of the changes in the expectation value of $T_{00}$ and these are in turn first order in the small parameter characterizing the spontaneous collapse and, as shown in \cite{ts}, are then successfully accounted for at first order in perturbation theory. The changes associated with the tensor modes are tied to changes in the expectation value of $T_{ij}$ and these are in turn second order in the small parameter characterizing the spontaneous collapse, and will thus appear only at second order in perturbation theory. 
 
%%%%%%%%%%%%%%%%%%%%%%%%%%%%%%%%%%%%%%%%%%%%%%%%%%%%%%%%%%%%%
\subsection{First order perturbations}
%%%%%%%%%%%%%%%%%%%%%%%%%%%%%%%%%%%%%%%%%%%%%%%%%%%%%%%%%%%%%

Here, we present a brief review of first order perturbation theory applied to inflation. We will focus only on results that will be useful for our approach (for a detailed analysis see \cite{mukhanov1992}). We are interested in studying the metric perturbations $\d g_{\mu\nu}$ and find it convenient to work in a specific gauge, the longitudinal one. Also, we shall focus on the scalar and  tensor perturbations only. Hence, the non-zero components of the perturbed metric are given by
\ba
g_{00}=-a^2(1+2\Phi),~~~~g_{ij}=a^2\[(1-2\Psi)\d_{ij}+h_{ij}\]~~~~(\text{where}~~ h_i^{~i}=0).
\ea
It follows that the non-zero components of the metric perturbations $ \d g_{\mu\nu}=g_{\mu\nu}-\bg_{\mu\nu}$ are given by
 \ba
\d g_{00}=-2\Phi a^2,~~~~\d g_{ij}=a^2(-2\Psi\d_{ij}+h_{ij}),\label{frwper}
\ea
from which one can also find the non-zero components of the perturbed inverse metric using $\d g^{\mu\nu}=-\bar g^{\mu\r}\bar g^{\nu\s}\d g_{\r\s}$. 

Regarding the inflaton, we consider first order perturbations to the homogeneous component
\ba
\phi=\phi_0(\n)+\d\phi(x^{\mu}).
\ea
As $\p_{\a}\phi_0=\dot\phi_0\d^0_{\a}$, the equation of motion for the field $\d \phi$ is
\ba
\d\ddot\phi -\p^2(\d\phi)+2(\dot a/a)\d\dot\phi+a^2(\p^2V/\p\phi^2)\d\phi-(\dot\Phi+3\dot\Psi)
\dot\phi_0-2\Phi\[\ddot\phi_0+2(\dot a/a)\dot\phi_0)\]=0.
\ea

On the other hand, the Einstein equations at first order in the perturbations $\d G_{\mu \nu} = 8 \pi G \d T_{\mu \nu}$, serve to relate the metric perturbations with the inhomogeneities (at first order) of the scalar field. Also, at the linear order, the different types of perturbations (scalar, vector and tensor) decouple from each other. Additionally, if no anisotropic stress is present, $\Phi = \Psi$. From all this, for scalar perturbations we obtain (see \cite{pss})
\begin{equation}\label{mainmetricscalar1}
\nabla^2 \Psi = 4 \pi G \dot \phi_0 \d \dot\phi,
\end{equation}
and for the tensor perturbations we get (see \cite{mukhanov1992,wein})
\begin{equation}\label{evo}
 \ddot h_{ij} + 2 \mH \dot h_{ij} - \nabla^2 h_{ij} =0.
\end{equation}

As discussed above, we take the conditions associated with early stages of inflation to correspond to a space-time exactly described by the spatially flat FRW solution. Therefore, the initial conditions must be taken as $ h_{ij} = \dot h_{ij} =0 $ at early times and, thus, the solution of the equation above is $h_{ij} =0$ at all times. 

As we already noted, and as we will next see in detail, things will change when we consider the next order in perturbation theory. 

%%%%%%%%%%%%%%%%%%%%%%%%%%%%%%%%%%%%%%%%%%%%%%%%%%%%%%%%%%%%%
\subsection{Second order perturbations}
%%%%%%%%%%%%%%%%%%%%%%%%%%%%%%%%%%%%%%%%%%%%%%%%%%%%%%%%%%%%%

Given that the source of primordial gravitational waves, within the collapse proposal, is at second order in matter fields, we need to focus on second order perturbations of the metric. The second order cosmological perturbation theory has been developed before (see \cite{acquaviva} for a detailed analysis). Choosing the generalized longitudinal gauge, the components of the perturbed metric up to second order are
\begin{equation}
\label{metric2nd}
 g_{00} = -a^2 [1+2\Psi^{(1)}+\Psi^\2], \quad g_{0i} = 0, \quad
 g_{ij} = a^2 \left[ (1-2\Psi^\1 - \Psi^\2 ) \delta_{ij} + \frac{1}{2} h_{ij}^\2 \right] ,
 \end{equation} 
 where $\Psi^{(1)}$, $\Phi^{(1)}$, $\Psi^{(2)}$ and $\Phi^{(2)}$ correspond to first an second order scalar perturbations. It is known that, at first order in the perturbative expansion, the amplitude of the vector modes decays rapidly during inflation \cite{mukhanovbook,bassett2005} and that, at second-order, vector modes can be produced via non-linear interaction of scalar (and tensor) modes \cite{lu2007}. Therefore, in the following we do not focus on vector modes. Note also that we are setting to zero the first order tensor modes; that is consistent with our approach, as described in the previous section, since first order scalar perturbations of matter fields do not act as source for the first order tensor perturbations. Finally, the inverse metric is obtained by requiring that $g^{ac} g_{cb} = \delta^{a}_{~b}$, up to second order in the perturbations.

For the sake of completeness, we also present the equation of motion at second order in the perturbations of the field using the longitudinal gauge of Eq. \eqref{metric2nd}. That is, we expand the scalar field up to second order,
\begin{equation}\label{field2order}
 \phi=\phi_0(\n)+\deu\phi(x^{\mu}) + \frac{1}{2} \ded\phi(x^\mu),
\end{equation} 
and then perturb the Klein-Gordon equation at second order. Additionally, to simplify some terms, we use the zeroth and first order equations of motion and the fact that $\Psi^{(1)} = \Phi^{(1)}$. The result is
\bea
\label{KG2L}
& & \frac{1}{2}{\ded \ddot \phi} \,+\,\Aa {\ded \dot
\phi} \,-\frac{1}{2} \La \ded\phi \,-\, {\Phi^{(2)}}\,{\ddot \phi_0}
\,-\, 2\,\Aa {\Phi^{(2)}}\,{\dot \phi_0} \,-\,
\frac{1}{2}\,{\dot \Phi^{(2)}}
{\dot \phi_0}
\\ &-&\, \frac{3}{2}\,{\dot \Psi^{(2)}}\,{\dot \phi_0}
\,-\, 4\,{\Psi^{(1)}}\,{\dot \Psi^{(1)}}{\dot \phi_0}
\,-\,4\, {\dot \Psi^{(1)}}{\deu \dot \phi} \,-\, 4\,{\Psi^{(1)}}\,\La \deu \phi \,= \nonumber \\
&-2& {\Psi^{(1)}}\,\deu{\phi}\frac{\partial^2 V}{\partial \phi^2}\,a^2\,-\,
\frac{1}{2}\ded \phi\, \frac{\partial^2 V}{\partial \phi^2}\,a^2
\,-\, \frac{1}{2}\,(\deu \phi)^2\,\frac{\partial^3 V}{\partial \phi^3} \,a^2 \, . \nonumber
\eea

As we mentioned previously, cosmological perturbation theory at second order has been studied in \cite{ananda2006,baumann2007,acquaviva}. In particular, the Einstein second order perturbed equations, $\delta^{(2)}G^i_{~j}={8 \pi G} \delta^{(2)} T^{i}_{~j}/2$, yield
\bea
& & \Bigg(\frac{1}{2}\,\LA \Phi^{(2)} \,+\, \Aa {\dot \Phi^{(2)}}
\,+\, \Ac \Phi^{(2)} \,+\, \Ab \Phi^{(2)} \,-\, \frac{1}{2}\,\LA \Psi^{(2)} \,+\, 2\,\Aa
{\dot \Psi^{(2)}} 
+ {\ddot \Psi^{(2)}} \label{Gijmain}\\
&-&8\,\Ac \left( \Psi^{(1)} \right)^2 \,+\, 4\,\Ab \left( \Psi^{(1)} \right)^2 
\,-\, 8\,\Aa \Psi^{(1)}{\dot \Psi^{(1)}}
\,-\,3\,\partial_k \Psi^{(1)}\,\partial^k \Psi^{(1)} \, -\, 4
\,\Psi^{(1)}\,\LA \Psi^{(1)} \,\nonumber \\
&-&\, \left( {\dot \Psi^{(1)}} \right)^2 \Bigg)\,\delta^i_{~j}\,-\frac{1}{2}\,\partial^i\,
\partial_j\,\Phi^{(2)} \,+\,
\frac{1}{2}\,\partial^i\partial_j \,\Psi^{(2)} \,+\, \frac{1}{2}\,\Aa {\dot h^{i(2)}_{~j}} \nonumber \\
&+& \frac{1}{4}\, {\ddot h^{i(2)}_{~j}}
\,-\, \frac{1}{4}\,\LA h^{i(2)}_{~j} \,+\, \,\partial^i \Psi^{(1)}\,\partial_j
\Psi^{(1)} \,+\, 2\,\Psi^{(1)}\,\partial^i\partial_j \Psi^{(1)}\nonumber \\
&=& 8 \pi G \Big(\frac{1}{2}\, {\ded\dot \phi}\,\dot \phi_0
\,-\, \frac{1}{2}\, {\ded\phi}\,\frac{\partial V }{\partial \phi}\,a^2 \,+\,
\frac{1}{2}\,\left( {\deu\dot \phi} \right)^2 \,-\, \frac{1}{2}\,\partial_k
\,{\deu\phi}\,\partial^k {\deu\phi} \,+\, 2\,\left( \Psi^{(1)} \right)^2\,
{\dot \phi_0}^2 \nonumber \\
&-& \frac{1}{2}\,\left( \deu\phi \right)^2 \,\frac{\partial^2 V}{\partial
\phi^2}\,a^2 \,-\, 2\,\Psi^{(1)}\,{{\deu \dot \phi}}
\,\dot \phi_0\Big)\,\delta^i_{~j}+\,\,\frac{8 \pi G}{2} \left(\partial^i
{\deu\phi} \,\partial_j {\deu\phi}\right) \nonumber \,.
\eea

Eqs. \eqref{mainmetricscalar1}, \eqref{evo} and \eqref{Gijmain} will be most useful for us in what follows.

%[ I THINK WE NEED TO SAY SOMTHING ABOUT the terms in $ \chi$]

%%%%%%%%%%%%%%%%%%%%%%%%%%%%%%%%%%%%%%%%%%%%%%%%%%%%%%%%%%%%%
%%%%%%%%%%%%%%%%%%%%%%%%%%%%%%%%%%%%%%%%%%%%%%%%%%%%%%%%%%%%
%\section{Primordial gravitational waves}\label{GW}
\section{The tensor power spectrum within the objective collapse framework}
\label{secPStensor}
%%%%%%%%%%%%%%%%%%%%%%%%%%%%%%%%%%%%%%%%%%%%%%%%%%%%%%%%%%%%%
%%%%%%%%%%%%%%%%%%%%%%%%%%%%%%%%%%%%%%%%%%%%%%%%%%%%%%%%%%%%%

Equation \eqref{evo} shows how, at first order, tensor perturbations of the metric do not have a matter field source. Therefore, the semi-classical gravity approach implies that $h^{(1)}_{ij}=0$, i.e., that there are no primordial gravitational waves at first order. As a consequence, we need to considered second order cosmological perturbation theory. Eq. \eqref{Gijmain} describes the relation between metric perturbations and perturbations associated with the inflaton, at first and second orders. Given that we are interested in waves characterized by $h^{\2}_{ij}$, which corresponds to a symmetric, transverse and traceless tensor, we can construct a projection tensor $\mP_{ij}^{~~lm}$ that extracts the transverse, traceless part of any tensor (see \cite{ananda2006,baumann2007} and the appendix \ref{AppPtensor}). Applying the projection tensor $\mP_{ij}^{~~lm}$ on both sides of Eq. \eqref{Gijmain} eliminates the contribution from the diagonal terms and from the objects $\Psi^\2$ and $\Phi^\2$. Therefore, the equation of motion corresponding to $h^{\2}_{ij}$ is given by
\ba
(-\p_0^2+\p^2) h_{ij}^{\2}(\v x,\n)-\f{2\dot a}{a}\dot h_{ij}^{\2}(\v x,\n)=S^{TT}_{ij}(\v x,\eta),\label{eom}
\ea
where $S^{TT}_{ij}(\v x,\eta)$ is the transverse and traceless part of
\ba
S_{ij}(\v x,\eta)=2\[4\Psi^{\1}(\v x,\n)\p_i\p_j\Psi^{\1}(\v x,\n)+2(\p_i\Psi^{\1}(\v x,\n))(\p_j\Psi^{\1}(\v x,\n))-8\pi G \{\p_i\d^{\1}\phi(\v x,\n)\} \{\p_j\d^{\1}\phi(\v x,\n)\}\].
\ea
From Eq. \eqref{eom}, we observe that the source of the second order tensor perturbations $h_{ij}^\2$ is given in terms of products of first order scalar perturbations, associated to both the metric and the inflaton. From this point on, we will omit the index $^\1$ from first order scalar perturbations and, as our object of interest is the second order tensor perturbation (given that the first order vanishes), we will also omit the index $^\2$ from $h_{ij}^\2$.

Considering the problem in a fiduciary box of side L, and passing to a description in terms of a Fourier decomposition, Eq. \eqref{eom} becomes, 
\ba
(\p_0^2+q^2) h_{ij}(\v q,\n)+\f{2\dot a}{a}\dot h_{ij}(\v q,\n)=\tilde S^{TT}_{ij}(\v q,\eta),
\ea
where $\tilde S^{TT}_{ij}(\v q,\eta)$ is the transverse and traceless part of 
\ba
\tilde S_{ij}(\v q,\eta)=\f{1}{L^3}\sum_{\v p}\[\{8p_ip_j+4(\v q-\v p)_ip_j\}\Psi(\v q-\v p, \n)\Psi(\v p,\n)-16\pi G(\v q-\v p)_ip_j\d\phi(\v q-\v p)\d\phi(\v p)\].
\ea
Now, without loss of generality, we choose $\v q=q\hat z$, i.e., $q_1=q_2=0, q_3=q$ and consider the component $i=1,j=2$ of the equation. Thus, we have
\ba
(\p_0^2+q^2) h_{12}(\v q,\n)+\f{2\dot a}{a}\dot h_{12}(\v q,\n)=\f{1}{L^3}\sum_{\v p}\[4p_1p_2\Psi(\v q-\v p, \n)\Psi(\v p,\n)+16\pi G p_1p_2\d\phi(\v q-\v p)\d\phi(\v p)\].~~~\label{eom12}
\ea

We recall that the scalar metric perturbation at first order is related to the inhomogeneous field $\d \phi$ as given by Eq. \eqref{mainmetricscalar1}. If we redefine the field as $y=a \d\phi$, and the corresponding conjugate momentum $\pi=\dot y-y\dot a/a$, then Eq. \eqref{mainmetricscalar1}, in discrete Fourier space, becomes 
\ba\label{masterescalar}
\Psi(\vq,\n)=-\f{4\pi G \dot \phi_0(\n)}{a(\n)}\f{\bra \hat \pi(\vq,\n) \ket}{q^2}.
\ea
Using the former expression and the rescaled field $y$, and turning to the semi-classical version, in view of incorporating the collapse dynamics, Eq. \eqref{eom12} becomes, 
\ba 
\(\f{\p^2}{\p\n^2}+q^2-\f{2}{\n}\f{\p}{\p\n}\) h_{12}(\v q,\n)=\f{16\pi G}{a^2L^3}\sum_{\v p}p_1p_2\[(4\pi G\dot\phi_0^2)\f{\langle\hat\pi(\v q-\v p)\rangle}{|\v q-\v p|^2}\f{\langle\hat\pi(\v p)\rangle}{p^2}+\langle \hat y(\v q-\v p)\rangle\langle\hat y(\v p)\rangle\],~~~
\ea
where we have replaced $\p_0$ by $\p/\p\n$ and used $\dot a/a=-1/\n$. Dropping the indices $1,2$, and using $a=-1/H\n$, we write the solution of the above differential equation as
\ba
h(\v q,\n)=-ih^+(\v q,\n)\int_{-T}^{\n} \f{h^-(\v q,\n') S(\v q,\n')}{H^2\n'^2}d\n'+ih^-(\v q,\n)\int_{-T}^{\n} \f{h^+(\v q,\n') S(\v q,\n')}{H^2\n'^2}d\n'-c_1h^+(\v q,\n)+c_2h^-(\v q,\n),
\ea
where 
\ba
h^{\pm}(\v q,\n)=-\f{H}{\sqrt{2q}}\(\n\pm\f{i}{q}\)e^{\pm iq\n}\label{defhpm}
\ea
and 
\ba\label{defsource}
S(\v q,\n) =\f{16\pi GH^2\n^2}{L^3}\sum_{\v p}p_1p_2\[4\pi G\dot\phi_0^2(\n)\f{\langle\hat\pi(\v q-\v p,\n)\rangle}{|\v q-\v p|^2}\f{\langle\hat\pi(\v p,\n)\rangle}{p^2}+\langle \hat y(\v q-\v p,\n)\rangle\langle\hat y(\v p,\n)\rangle\].
\ea
Since at the beginning of inflation, $\n=-T$, we must have $h(\v q,-T)=0$, the solution reduces to
\ba
h(\v q,\n)=-ih^+(\v q,\n)\int_{-T}^{\n} \f{h^-(\v q,\n') S(\v q,\n')}{H^2\n'^2}d\n'+ih^-(\v q,\n)\int_{-T}^{\n} \f{h^+(\v q,\n') S(\v q,\n')}{H^2\n'^2}d\n' . \label{hqn}
\ea
Note that $h(\v x,\n)$ is dimensionless, which implies its Fourier transform $h(\v q,\n)$ has dimension $L^3$ in the units we are using. Eq. \eqref{hqn} is the main result we will be using below to compute the tensor spectrum within our scheme. We can see that primordial gravitational waves are sourced by quantum expectation values of matter fields at second order in the perturbations. These quantum expectation values are zero for the initial Bunch-Davies vacuum. It is only after it undergoes a spontaneous collapse that it acquires a non-zero value. Only at such point the produced matter and curvature perturbations give rise to primordial gravitational waves.

%\section{The tensor power spectrum within the objective collapse framework}\label{secPStensor}

In what follows we show in detail how the standard prediction for the tensor power spectrum is modified within the context of a semi-classical treatment, augmented with the spontaneous collapse hypothesis. First we employ a CSL model \cite{per,bassi,bassi2012} where a modification of the Schr\"odinger equation leads naturally to the eventual collapse of the inflaton wave function \cite{ppd,LB15}. We then compare the results of the CSL model with those obtained with a simpler collapse model, which we call the Newtonian collapse scheme \cite{adolfo2008,claudia,pia,micol2}, which is based on a phenomenological parametrization of the post-collapse state.\footnote{ A similar analysis, with a very particular choice of the model's parameters, was done in \cite{glsus}; see also \cite{mauro,qmatter} for a treatment of the primordial tensor modes, using the collapse proposal, within the traditional framework of quantizing both the metric and the matter fields during inflation and in a bouncing cosmological model respectively.} Our motivation for this comparison is to explore the robustness of the results. Before discussing each of these models separately, though, we need to explain how the averages required to calculate the power spectrum are to be understood and computed.

The standard definition of the tensor power spectrum $P_h$ is given by
\begin{equation}\label{PSoriginal}
 \avg{h(\vq_1,\eta) h(\vq_2,\eta)^*} = {2\pi^2} P_h (q_1,\eta) \delta(\vq_1-\vq_2),
\end{equation}
with $\avg{\cdot}$ denoting an average over possible realizations of $h(\vq,\eta)$. In the traditional inflationary paradigm, one makes the (unwarranted) identification $\bra 0 | \hat h(\vq_1,\eta) \hat h(\vq_2,\eta)^\dag | 0 \ket = \avg{h(\vq_1,\eta) h(\vq_2,\eta)^*} $. However, within our scheme, the object $h(\vq,\eta)$ acquires a stochastic character inherited from the collapse of the quantum state of the matter fields. Therefore, within our approach, the average in Eq. \eqref{PSoriginal}, is computed over possible realizations of the stochastic function involved. The quantity $\avg{h(\vq_1,\eta) h(\vq_2,\eta)^*}$ is also needed in order to obtain the expression for observable quantities, such as the $C_l$'s for the B-modes of the polarization of the CMB. We can always relate the aforementioned average with the most likely value of the observables in exactly the same way as was done in \cite{ppd}. Therefore, once the value of \eqref{PSoriginal} is obtained, physical observables are straightforwardly computed.

Hence, our next task is to compute $\avg{h(\vq_1,\eta) h(\vq_2,\eta)^*}$. Here, we consider that even though actual measurements of B-modes in the CMB are associated with the power spectrum at the time of decoupling, we evaluate the tensor spectrum at the end of the inflationary era, i.e., when $\eta \to 0^-$. We believe this is warranted because, it is very hard to conceive of a physical process during the radiation epoch that could amplify the power spectrum in a substantial manner. In the limit $\eta \to 0^-$, from Eq. \eqref{hqn} (using trigonometrical properties and the definition of Bessel function of order 3/2) we obtain 
\begin{equation}\label{mainPStensor}
 \avg{h(\vec q_1,0^-) h(\vec q_2,0^-)^*} = \frac{\pi}{2} \frac{1}{q_1^2} \int_{-T}^{0^-} d\eta_1 \int_{-T}^{0^-} d\eta_2 \frac{J_{3/2} (q_1 \eta_1)}{\sqrt{q_1 \eta_1}} \frac{J_{3/2} (q_1 \eta_2)}{\sqrt{q_1 \eta_2}} \: \avg{S(\vec q_1,\eta_1)S^*(\vec q_2,\eta_2)}.
\end{equation}

As can be seen from Eq. \eqref{mainPStensor}, the information regarding the collapse process is contained in the object $\avg{S(\vec q_1,\eta_1)S^*(\vec q_2,\eta_2)}$, so let us focus on that quantity. In order to make the source $S(\vec q,\eta)$ more tractable, we define the vector $\vec \kappa \equiv \vec q - \vec p$, which implies $ \kappa \equiv |\vec q - \vec p |$. Also, using the slow-roll approximation and the definition of the reduced Planck's mass 
$M_P^2 \equiv 1/(8\pi G )$, we have $4 \pi G \dot \phi_0^2 = \frac{\epsilon}{\eta^2}$, where $\epsilon$ is the slow-roll parameter. Therefore, the source term $S(\vec q,\eta)$, Eq. \eqref{defsource}, can be rewritten in terms of the slow-roll parameter $\epsilon$ which is very small. As a consequence, at this point, it is convenient to neglect all the terms proportional to the slow-roll parameter in the source term. Thus, the ensemble average of the source terms is approximately given by 
\begin{equation}\label{productosS2app}
\avg{S(\vec q_1,\eta_1)S^*(\vec q_2,\eta_2)} \simeq \frac{4H^4}{M_P^4 L^6} 
 \sum_{\vec p, \vec p\:'} p_1 p_2 p_1' p_2' \bigg[ \eta_1^2 \eta_2^2 \avg{\expec{\hat y (\vec \kappa_1,\eta_1)} 
\expec{\hat y(\vec p,\eta_1) } \expec{\hat y (\vec \kappa_2,\eta_2) }^* \expec{\hat 
y (\vec p\:', \eta_2)}^* }
\bigg].
\end{equation}
The rest of the calculations are straightforward. Use the the CSL inflationary model or the Newtonian collapse scheme to compute the average in Eq. \eqref{productosS2app}, then use the resulting average to obtain the tensor power spectrum from Eq. \eqref{mainPStensor}.
\subsection{The tensor power spectrum in the CSL inflationary model}
\label{secPSCSL}
%%%%%%%%%%%%%%%%%%%%%%%%%%%%%%%%%%%%%%%%%%%%%%%%%%%%%%%%%%%%%

In this subsection we will employ the CSL inflationary model first developed in \cite{ppd} (observational tests for such a model were recently explored in \cite{micol2}). Generically the (non-cosmological) CSL model is based on a modification of the Schr\"odinger equation that induces a collapse of the wave function unto one of the eigenstates of a, so-called, collapse operator. The collapse process is induced by the interaction of the system with a background noise $\Omega(t)$, which is a continuous stochastic process of the Wiener kind. Note that the noise $\Omega(t)$ is characterized only through its statistical properties (so it is not a parameter of the theory). The rate of collapse is controlled by the CSL parameter $\lambda$ (see \cite{bassi,bassi2012} for a thorough review).

In order to apply the CSL model to the inflationary setting, we will follow the approach first introduced in \cite{ppd}. That work relies on a version of the CSL model in which its nonlinear aspects are shifted to the probability law. That is, the evolution law is linear just as the Schr\"odinger equation, but then, the law of probability for the realization of a specific random function, becomes dependent of the state that results from such evolution. Specifically, the theory can be characterized in terms of two equations. The first is a modified Schr\"odinger equation, whose solution is
\begin{equation}\label{cslevol}
|\psi,t\rangle={\cal T}e^{-\int_{0}^{t}dt'\big[i\hat H+\frac{1}{4\lambda}[w(t')-2\lambda\hat A]^{2}\big]}|\psi,0\rangle.
\end{equation}
$\cal T$ is the time-ordering operator, $w(t)$ is a random classical function of time of white noise type. The probability for this $w(t)$ is given by the second equation, the Probability Rule
\begin{equation}
	PDw(t)\equiv\langle\psi,t|\psi,t\rangle\prod_{t_{i}=0}^{t}\frac{dw(t_{i})}{\sqrt{ 2\pi\lambda/dt}}.
\end{equation}

In the case of multiple identical particles in 3 dimensions, the CSL theory would contain one stochastic function for each independent degree of freedom, but only one parameter $\lambda$. In the case of several species of particles,
 the theory would naturally involve a parameter $\lambda_i$ for each particle species (one might postulate for them to be equal, but that is not necessary). In fact, there is strong phenomenological preference for a $\lambda_i$ that depends on the particle’s mass $m_i$ (see \cite{gpr,per}).

Given that the CSL model modifies the Schr\"odinger equation, it is convenient to describe the quantum theory of the inflaton in the Schr\"odinger picture, where the relevant objects are the wave function and the Hamiltonian. The Hamiltonian characterizing the inhomogeneous sector of the inflaton is $H = (1/2) \int d^3q (H_{\vq}^\textrm{R} + H_{\vq}^\textrm{I})$ with
\begin{equation}
 H_{\vq}^\RI = \pi_{\vq}^\RI \pi_{\vq}^{*\RI} + q^2 y_{\vq}^\RI y_{\vq}^{*\RI} - \frac{\dot a}{a} \left( \yk^\RI \pk^{*\RI} + \yk^{*\RI} \pk^{\RI} \right)
\end{equation} 
where $\yk = a \dphi_{\vq}$ and $\pi_{\vq} \equiv y_{\vq} '- \mH y_{\vq} $. The indexes R,I denote the real and imaginary parts of $\yk$ and $\pk$. We now promote $\yk$ and $\pk$ to quantum operators by imposing canonical commutations relations $[\hat y_{\vq}^\RI, \hat \pi_{\vq}^\RI] = i \delta (\vq-\vq')$. The CSL inflationary model is based on the assumption that the objective collapse mechanism acts on each mode of the field independently. Therefore, generalizing equation \eqref{cslevol}, 
 the evolution of the state vector corresponding to the state of the field is
\begin{equation}\label{CSLevolution}
 |\Phi_{\vq}^{\textrm{R,I}}, \eta \ket ={ \cal T} \exp \bigg\{ - \int_{-T}^{\eta} 
d\eta' \bigg[ i \hat{H}_{\vq}^{\textrm{R,I}} + \frac{1}{4 \lambda_q} (\Omega(\vq,\eta') - 2 \lambda_q^2 
\hat{\pi}_{\vq}^{\textrm{R,I}})^2 \bigg] \bigg\} |\Phi_{\vq}^{\textrm{R,I}}, 
 \tau \ket,
\end{equation} 
where $\cal T$ is the time-ordering operator and $-T$ denotes the conformal time at the beginning of inflation.

Since we are applying the CSL collapse dynamics to each mode of the field, it is natural to introduce a stochastic function for each independent degree of freedom. That is, $\Omega$ should depend on $\vq$. Consequently, $\Omega_q$ is a stochastic field, which might be regarded as a Fourier transform on a stochastic space-time field $\Omega (x, t)$ (see a more detailed discussion in \cite{ppd}). The statistical properties of the field $\Omega_q$ are given in Eq. \eqref{omegasabhi} (where $\omega_{\beta}$ with $\beta=$R,I correspond to the the real and imaginary parts of the stochastic function $\Omega_q$, and thus are naturally dependent on $q$).
 
In \cite{ppd}, the possibility for an effective dependence of the parameter $\lambda$ on $q$ was uncovered. As shown there, such possility must be viewed as resulting from the specific form the collapse operator has, as expressed in terms of field variables. In the case at hand, as discussed in that work, it must be taken to be a suitable derivative of the the momentum conjugate to the inflaton field. When passing to the Fourier decomposition, the choice for the collapse operator translates into what seems as a dependence of $\lambda$ on $q$. The choice of the momentum operator $\hat{\pi}_{\vq}^{\textrm{R,I}}$ as the collapse operator is obtained by taking derivatives of $\hat{\pi}_{\vq}^{\textrm{R,I}}$, which  is motivated by the fact that the metric perturbation $\Psi$ is directly related to the expectation value of the momentum operator (see Eq. \ref{masterescalar}). For more details we ask the reader to consult \cite{ppd}.

Given that $\hat{\pi}_{\vq}^{\textrm{R,I}}$ has been chosen as the collapse operator, it is easier to work with a wave function in the momentum representation. In Fourier space, the wave function can be factorized into mode components $\Phi[\pi] = \Pi_{\vq} \Phi_{\vq}^\textrm{R} [\pi_{\vq}^\textrm{R}] \times \Phi_{\vq}^\textrm{I} [\pi_{\vq}^\textrm{I}] $. We consider the wave function of each mode of the field to be a Gaussian during the whole evolution; specifically
\begin{equation}\label{wf}
 \Phi^\RI (\eta,\pk^\RI) = \exp[-A_q(\eta) (\pk^\RI )^2 + B_q^\RI(\eta) \pk^\RI + C_q^\RI (\eta) ].
\end{equation} 
The set of equations describing the system are thus the dynamical equations for the objects $A_q(\eta)$, $B_q^\RI(\eta)$, and $C_q^\RI (\eta)$. These equations are found by inserting the wave function depicted in Eq. \eqref{wf} into the CSL evolution equation \eqref{CSLevolution}. The initial conditions are set by the initial state of the field, i.e., the Bunch-Davies vacuum. That is, the initial conditions are $A_q(-T) = 1/2q$, $B_q^\RI(-T)=0$, and $C_q^\RI (-T)=0$. Given that we are mainly interested in the expectation value $\bra \hat{y} (\vq,\eta) \ket$, only the solution $A_q(\eta)$ will be of importance. The evolution equation of $A_q(\eta)$ is 
\begin{equation}
\dot{A}_q(\eta) = \frac{i}{2} + \lambda_q - \frac{2}{\eta} A_q(\eta) - 2ik^2 A_q(\eta)^2,
\end{equation}
whose solution with the initial condition is
\begin{equation}\label{Ak}
A_q(\eta) = \frac{i}{2q^2 \eta} + \frac{\alpha_q}{2iq^2} \left[ \frac{(1-iqT) \cos \alpha_q (\eta + T) + \alpha_q T \sin \alpha_q (\eta + T) }{(1-iqT) \sin \alpha_q (\eta + T) - \alpha_q T \sin \alpha_q (\eta + T) } \right]
\end{equation}
with 
\begin{equation}
\alpha_q \equiv q \sqrt{1-2i\lambda_q}. 
\end{equation}
Hence, the quantity $A_q (\eta)$ depends on the CSL parameter $\lambda_q$ (through $\alpha_q $).

Using the wave function \eqref{wf}, which follows the evolution equation \eqref{CSLevolution}, the quantum expectation value of the field $\hat y(\vq,\eta) $ can be calculated in terms of the noise function
\begin{equation}\label{expecy}
 \expec{\hat y (\vec q,\eta)} = \frac{i L^{3/2}}{2^{3/2} q^2 (A_q(\eta) +A_q^*(\eta) )} \int_{-T}^\eta 
 d\eta' \Omega(\vq,\eta') F_{q} (\eta,\eta'),
\end{equation} 
with 
\begin{equation}\label{Fk}
F_{q} (\eta,\eta') \equiv \bigg[ \bigg( 
 \frac{-i}{\eta} + \alpha_q^* \bigg) 
e^{-i\alpha_q 
(\eta-\eta')} - \bigg( 
 \frac{i}{\eta} + \alpha_q \bigg) 
e^{i\alpha_q^* (\eta-\eta')} \bigg].
\end{equation} 
The function $F_{q} (\eta,\eta') $ depends on the CSL parameter $\lambda_q$ because of the $\alpha_q$.
Note that from Eq. \eqref{Ak}, and taking into account that $|\eta| \ll T$, we have that $(\textrm{Re}[A_q(\eta)])^2 \simeq \frac{1+\sqrt{1+4\lambda_q^2}}{8q^2}$, which is independent of $\eta$. We can further split the noise function $\Omega(\vq,\eta)$ in its ``real'' and ``imaginary parts'' $\Omega(\vq,\eta) = w_R(\vq,\eta) + i w_I(\vq,\eta)$ (formally, they correspond to the symmetric and antisymmetric part of the noise $\Omega(\x,\eta)$). Given the expectation value in Eq. \eqref{expecy} we can calculate the average from Eq. \eqref{productosS2app}
\begin{eqnarray}\label{productosS3}
 & & \avg{S(\nq_1,\eta_1 S(\nq_2,\eta_2)^* }= \frac{H^4}{2^8 M_P^4 L^6} \sum_{\vp, 
\vp'} p_1 p_2 p_1' p_2' \frac{L^6}{ \textrm{Re}[A_{\kappa_1}] \textrm{Re}[A_{\kappa_2} 
] \textrm{Re}[A_{p}] \textrm{Re}[A_{p'} ]} \nn
&\times& \frac{1}{\kappa_1^2 \kappa_2^2 p^2 p'^2} \int_{-T}^{\eta_1} d\eta' 
\int_{-T}^{\eta_1} d\eta'' \int_{-T}^{\eta_2} d\eta''' \int_{-T}^{\eta_2} d\eta'''' \avg{ 
 \Omega(\vkappa_1,\eta') \Omega(\vp,\eta'') \Omega(\vkappa_2,\eta''')^* \Omega 
(\vp',\eta'''')^* } \nn
&\times& \bigg[ \eta_1^2 \eta_2^2 
F_{\kappa_1} (\eta_1,\eta') F_{p} (\eta_1,\eta'') 
F_{\kappa_2} (\eta_2,\eta''')^* F_{p'}(\eta_2,\eta'''')^* 
\bigg].
\end{eqnarray}
In Appendix \ref{app1} we present the details of the calculations of the average $\avg{S(\nq_1,\eta_1 S(\nq_2,\eta_2)^* }$. The resulting expression yields
\begin{equation}\label{productosS4}
\avg{S(\nq_1,\eta_1 S(\nq_2,\eta_2)^* } = \frac{H^4}{2^7 M_P^4} \sum_{\vp} 
\frac{p_1^2 p^2_2 (\delta_{\vq_1,\vq_2} + \delta_{\vq_1,-\vq_2})}{p^4 
\kappa_1^4 (\textrm{Re}[A_p])^2 (\textrm{Re}[A_{\kappa_1}])^2} \eta_1^2 \eta_2^2 I_{FF} 
(p;\eta_1,\eta_2) I_{FF} (\kappa_1;\eta_1,\eta_2) ,
\end{equation}
where we have defined
\begin{equation}\label{defIXX}
 I_{FF} (k;\eta_1,\eta_2) \equiv \int_{-T}^{\eta_1} d\eta' \int_{-T}^{\eta_2} d\eta'' 
 K(k,k;\eta',\eta'') F_k (\eta_1,\eta') F_k^* (\eta_2,\eta''),
\end{equation} 
and the function $K(k,k;\eta',\eta'')$ is defined in Eq. \eqref{K}. The explicit solution of the integral $I_{FF}$ is a very cumbersome procedure; therefore the interested reader can consult the specific details in Appendix \ref{app1} (see Eqs. \eqref{defIFF}-\eqref{IFF}). In the ensuing expressions we will just leave indicated the integral $I_{FF}$, which depends on the CSL parameter $\lambda_q$ because of the $F_{q} (\eta,\eta')$ function. 

Finally, we can plug in Eq. \eqref{productosS4}, which is our final expression for $ \avg{S(\nq_1,\eta_1 S(\nq_2,\eta_2)^* }$, into Eq. \eqref{mainPStensor}. Note that the resulting expression for the average $\avg{S(\nq_1,\eta_1 S(\nq_2,\eta_2)^* 
 }$ establishes that Fourier modes such that $q_1 \equiv |\vq_1| \neq |\vq_2| \equiv q_2$ are uncorrelated (because of $\delta_{\vq_1,\vq_2}$, $\delta_{\vq_1,-\vq_2}$). As a result, we can make the replacement $q_2 \to q_1$ in Eq. \eqref{mainPStensor}. Thus, switching from discrete $\vp$ to the continuum, we arrive at
\begin{eqnarray}\label{PStensor4}
 \avg{h(\vec q_1,0^-) h(\vec q_2,0^-)^*} &=& \frac{H^4}{ q_1^3 2^8 M_P^4} 
 [\delta(\vq_1-\vq_2) + 
\delta(\vq_1+\vq_2)] q_1 \int d^3 p 
\frac{p_1^2 p^2_2 }{p^4 
\kappa_1^4 (\textrm{Re}[A_p])^2 (\textrm{Re}[A_{\kappa_1}])^2} \nn
&\times& \int_{-T}^0 d\eta_1 \int_{-T}^0 d\eta_2 \: \frac{1}{\sqrt{q_1 \eta_1 q_1 
 \eta_2}} \frac{\pi}{2} J_{3/2} (q_1 \eta_1) J_{3/2} (q_1 \eta_2) \nn
&\times& \eta_1^2 \eta_2^2 I_{FF} 
(p;\eta_1,\eta_2) I_{FF} (\kappa_1;\eta_1,\eta_2) .
\end{eqnarray} 

Next we extract the precise expression for the tensor power spectrum from Eq. \eqref{PStensor4}. By making the change of variables $\kappa_1 \equiv u q_1$, $p \equiv v q_1$, $x_1 \equiv q_1 \eta_1$ and $x_2 \equiv q_1 \eta_2$, Eq. \eqref{PStensor4} can be recast as 
 \begin{eqnarray}\label{PStensor5}
 \avg{h(\vec q_1,0^-) h(\vec q_2,0^-)^*} &=& \frac{ \pi^2 H^4 
\lambda_{q_1}^2}{ q_1 2^5 M_P^4} 
 [\delta(\vq_1-\vq_2) + 
\delta(\vq_1+\vq_2)] \int_0^{\infty} dv \int_{|1-v|}^{|1+v|} du 
\frac{[4v^2-(u^2-v^2-1)^2]^2}{uv 
 [u+(u^2+4\lambda_{q_1}^2)^{1/2}][v+(v^2+4\lambda_{q_1}^2 )^{1/2} ] } \nn
&\times& \int_{-q_1T}^0 dx_1 \int_{-q_1T}^0 dx_2 \: \frac{1}{\sqrt{x_1 x_2}} 
J_{3/2} (x_1) J_{3/2} (x_2) \nn
&\times& \frac{x_1^2 x_2^2 uv}{q_1^4 16 \lambda_q^2} I_{FF} ( vq_1; 
x_1/q_1,x_2/q_1) I_{FF} ( uq_1; x_1/q_1,x_2/q_1).
\end{eqnarray}

Finally, from \eqref{PStensor5} we can extract the power spectrum associated to the tensor modes
 \begin{eqnarray}\label{PStensor6}
 P_h (q_1) &=& \frac{ \pi^2 H^4 
\lambda_{q_1}^2}{ q_1 2^5 M_P^4} 
 \int_0^{\infty} dv \int_{|1-v|}^{|1+v|} du 
\frac{[4v^2-(u^2-v^2-1)^2]^2}{uv 
 [u+(u^2+4\lambda_{q_1}^2)^{1/2}][v+(v^2+4\lambda_{q_1}^2 )^{1/2} ] } \nn
&\times& \int_{-q_1T}^0 dx_1 \int_{-q_1T}^0 dx_2 \: \frac{1}{\sqrt{x_1 x_2}} 
J_{3/2} (x_1) J_{3/2} (x_2) \nn
&\times& \frac{x_1^2 x_2^2 uv}{q_1^4 16 \lambda_q^2} I_{FF} ( vq_1; 
x_1/q_1,x_2/q_1) I_{FF} ( uq_1; x_1/q_1,x_2/q_1).
\end{eqnarray} 

In Appendix \ref{app2}, we sketch our estimation of all the integrals involved in Eq. \eqref{PStensor6}. The result is
\begin{equation}\label{PStensor10}
 P_h (q_1) \simeq \frac{10^{-3} \pi H^4 \lambda_{q_1}^2}{ q_1^3 M_P^4} q_1^4 T^4 ,
v_m^5,
\end{equation} 
where we have introduced an UV cut-off $v_m \equiv p_{\textrm{UV}}/q_1$. The UV cut-off $v_m$ comes from the indefinite integral in Eq. \eqref{PStensor6}, otherwise we would have a divergent result. We will discuss this aspect at the end of the section. 

In Ref. \cite{ppd} the scalar power spectrum was deduced within the CSL inflationary model. The explicit expression for the dimensionless scalar spectrum is
\begin{equation}
 \mP_s (q) \simeq \frac{H^2 \lambda_q q T }{\epsilon M_P^2}.
\end{equation} 
Therefore, if $\lambda_q$ is independent of $q$ the resulting scalar spectrum would fail to be scale invariant, which is well known to be required for observational viability of the theory. In fact, scale invariance of the spectrum can only be achieved by including an the explicit dependence on $q$ of $\lambda_q$, of the form
\begin{equation}
\lambda_q = \frac{\lambda_0}{q},
\end{equation}
with $\lambda_0$ the CSL universal collapse parameter. As shown in \cite{ppd}, such dependence is both natural on dimensional grounds and can be achieved naturally by taking the collapse operator to be a suitable spatial derivative of the momentum conjugate to the filed. However as we will see, such $q$ dependence of $\lambda_q$ will result in a spectrum for tensor perturbations, Eq. \eqref{PStensor10}, which will not be scale invariant. In the next section we will discuss some of the observational consequence of that result.

% On the other hand, using the fact that the observational data puts a constraint to the amplitude of the spectrum to be $P_s (q)\sim 10^{-9}$, we have the follwing relation 
% \begin{equation}\label{key5}
% \frac{H^4 \lambda_q^2 q^2 T^2 }{M_P^4 } \simeq \epsilon^2 10^{-18}.
% \end{equation} 
% The previous relation is important because it allows to set an estimate for the amplitude of the tensor spectrum, that is, substituting \eqref{key5} into \eqref{PStensor10}, we find that 
% \begin{equation}
% P_h (q_1) \simeq \frac{10^{-21} \epsilon^2 }{q_1^3} q_1^2 T^2 v_m^5 = 
% \frac{10^{-21} \epsilon^2 }{q_1^6} T^2 p_{\textrm{UV}}^5.
%\end{equation} 

%%%%%%%%%%%%%%%%%%%%%%%%%%%%%%%%%%%%%%%%%%%%%%%%%%%%%%%%%%%%%
\subsection{The tensor power spectrum in the Newtonian collapse scheme}
%%%%%%%%%%%%%%%%%%%%%%%%%%%%%%%%%%%%%%%%%%%%%%%%%%%%%%%%%%%%%

With the purpose of analyzing the generality and robustness of the predictions for the tensor power spectrum within the self-induced collapse hypothesis, we present calculations for the tensor power spectrum based on what we call the Newtonian collapse scheme approach. The description of the collapse process in such an approach is essentially a very simple toy model, originally introduced in \cite{pss}. In this toy model, we assume that each mode $\vk$ undergoes a single collapse, which occurs at the conformal time $\tc$. Moreover, the characterization of the post-collapse state is given by the expectation values and the quantum uncertainties corresponding to the field and its conjugated momentum at the time $\tc$. 

In more detail, the expectation value of the field $\hat y(\nk,\eta) $ is generically taken to be given by (see \cite{Leon2010})
\begin{eqnarray}\label{expecynewt}
 \bra \hat y (\nk,\eta) \ket &=& \bigg[ \frac{\cos (k\eta - z)}{k}\bigg( \frac{1}{k \eta} - \frac{1}{z} \bigg) + \frac{\sin (k\eta - z) }{k} \bigg( \frac{1}{k\eta z} + 1 \bigg) \bigg] \bra \hat \pi (\nk, \tc) \ket \nn
 &+& \bigg[ \cos (k\eta - z)- \frac{(k\eta - z)}{k\eta} \bigg] \bra \hat y (\nk, \tc) \ket ,
\end{eqnarray}
where $z \equiv k \tc$. In order for the scalar power spectrum to be consistent with the observational data, in \cite{pss,claudia,micol,micol2} is found that the time of collapse has to be of the form $\tc \propto 1/k$, which implies that $z$ is independent of $k$. We also assume that the collapse affects only the conjugated momentum variable. Splitting into real and imaginary parts, the collapse can be characterized by 
\beq\label{esquemanewt}
 \bra \hat{y}^{R,I}_{\nk}(\eta^c_{\nk})\ket = 0, \qquad
 \bra \hat{\pi}^{R,I}_{\nk}(\tc) \ket = x_{\nk}^{R,I}
 \sqrt{\left(\Delta \hat{\pi}^{R,I}_{\nk} (\tc) \right)^2_0},
\end{equation}
where $x_{\nk}^{(R,I)}$ represents a random Gaussian variable normalized and centered at zero. The quantum uncertainty of the vacuum state, associated to the conjugated momentum at the time of collapse, is $\left(\Delta \hat{\pi}^{R,I}_{\nk} (\tc) \right)^2_0 = {L^{3} k }/{4}$.

Using Eqs. \eqref{expecynewt} and \eqref{esquemanewt} we can rewrite the expectation value of the field variable as
\begin{equation}\label{expecynewt2}
 \expec{\hat y(\vk,\eta)} = \Ny (\vk,\eta) \frac{L^{3/2}}{2} X_{\vk},
\end{equation}
where 
\begin{equation}
 \Ny (\vk,\eta) \equiv \bigg[ {\cos (k\eta - z)}\bigg( \frac{1}{k \eta} - \frac{1}{z} \bigg) + {\sin (k\eta - z) } \bigg( \frac{1}{k\eta z} + 1 \bigg) \bigg] \frac{1}{\sqrt{k}}
\end{equation}
and $X_{\vk} \equiv x_{\vk}^R + i x_{\vk}^I$. With Eq. \eqref{expecynewt2} at hand, we can proceed to determine the ensemble average of Eq. \eqref{productosS2app}
\begin{eqnarray}\label{expec4y}
 & & \avg{\expec{\hat y (\vec \kappa_1,\eta_1)} \expec{\hat y(\vec p,\eta_1) } \expec{\hat y (\vec \kappa_2,\eta_2) }^* \expec{\hat y (\vec p\:', \eta_2)}^*} = \nn
 & & \frac{L^6}{2^4} \Ny (\vkappa_1,\eta_1) \Ny (\vp,\eta_1) \Ny (\vkappa_2,\eta_2) \Ny (\vp',\eta_2) \avg{ X_{\vkappa_1} X_{\vp} X_{\vkappa_2}^* X_{\vp'}^* }.
\end{eqnarray}
As we have assumed that the random variables $X_{\vk}$ are Gaussian-distributed, we can write
\begin{equation}
 \avg{ X_{\vkappa_1} X_{\vp} X_{\vkappa_2}^* X_{\vp'}^* } = \avg{X_{\vkappa_1} X_{\vp} } \times \avg{ X_{\vkappa_2}^* X_{\vp'}^* } + \avg{X_{\vkappa_1} X_{\vp'}^* } \times \avg{ X_{\vkappa_2}^* X_{\vp} } + \avg{X_{\vkappa_1} X_{\vkappa_2}^* } \times \avg{ X_{\vp} X_{\vp'}^* }. 
\end{equation}
Moreover, $\avg{X_{\vk} X_{\vk'}^*} = 2 \delta_{\vk,\vk'}$ and $\avg{X_{\vk} X_{\vk'}} =\avg{X_{\vk}^* X_{\vk'}^*} = 2 \delta_{\vk,-\vk'}$, so the ensemble average is given by
\begin{equation}\label{expec4x}
 \avg{ X_{\vkappa_1} X_{\vp} X_{\vkappa_2}^* X_{\vp'}^* } =2 [ \delta_{\vkappa_1,-\vp} \delta_{\vkappa_2,-\vp'} + \delta_{\vkappa_1,\vkappa_2} \delta_{\vp,\vp'} + \delta_{\vkappa_1,\vp'} \delta_{\vkappa_2,\vp}].
\end{equation}

Substituting Eqs. \eqref{expec4y} and \eqref{expec4x} in Eq. \eqref{productosS2app}, and performing the sum over $\vp'$, and passing to the continuum limit ( $L \to \infty$) we find, 
\begin{equation}\label{productosS2app2}
 \avg{S(\vec q_1,\eta_1)S^*(\vec q_2,\eta_2)} \simeq \frac{4H^4}{M_P^4} \frac{\delta(\vq_1- \vq_2)}{2^2} 
\int d^3 p \quad p_1^2 p_2^2 \eta_1^2 \eta_2^2 \Ny(|\vq_1-\vp|,\eta_1) \Ny(p,\eta_1) \Ny(|\vq_1-\vp|,\eta_2) \Ny(p,\eta_2),
\end{equation}
where we have also used the definition $\vec \kappa_1 \equiv \vq_1 - \vp$.

% From the former expression, we can estimate the degree of divergence in $p$ by power counting. The factors $d^3 p$ $ p_1^2 p_2^2$ yield a scaling of $p^7$; on the other hand, the four products of the $\Ny(\vk,\eta)$ functions scale as $\sim (1/p^6 + 1/p^2)$. Therefore, the divergence in $p$ scales as $p + p^5$, evidently, the dominant term is $p^5$. We will continue with the calculations in order to corroborate our estimation. 

Our next task is to use Eq. \eqref{mainPStensor}, together with Eq. \eqref{productosS2app2}, to estimate the amplitude of the tensor power spectrum. As in the case of the CSL inflationary approach, we choose to evaluate the spectrum at the end of the inflationary regime, i.e., $\eta \to 0^-$. Consequently, Eq. \eqref{mainPStensor} is explicitly given by
\begin{equation}\label{PStensor11}
 \avg{h(q_1,0^-) h(q_2,0^-)} = \frac{H^4}{M_P^4 q_1^6} \frac{\pi}{2} \delta(\vq_1-\vq_2) \int d^3 p \quad p_1^2 p_2^2 I_{NN}^2,
\end{equation}
where
\begin{equation}
 I_{NN} \equiv \int_{\eta_{q_1}^c}^{0^-} d\eta_1 \ (q_1 \eta_1)^{3/2} J_{3/2} (q_1 \eta_1) \Ny(|\vq_1-\vp|,\eta_1) \Ny(p,\eta_1).
\end{equation}

It is important to note that the integral $I_{NN}$ begins at $\eta_{q_1}^c$. Physically, this means that the expectation value of the field variable acts as source for the tensor modes only after the time of collapse $\eta_{q_1}^c$. Additionally, it is convenient to perform a change of variable $x = q_1 \eta_1$. This implies that the lower limit of integration is changed to $q_1 \eta_{q_1}^c$. However, since $\eta_{q_1}^c \propto 1/q_1$, the lower limit of integration is simply $z$, with $z$ independent of $q_1$.

In order to provide an estimate for $I_{NN}$, we can neglect the oscillatory factors in the functions $\Ny$, hence
\begin{equation}
\Ny(k,\eta) \simeq \frac{A_k}{q_1 \eta} + B_k, 
\end{equation}
where
\begin{equation}
 A_k \equiv \frac{1}{k^{3/2}} \left( 1 + \frac{1}{z} \right), \qquad B_k \equiv \frac{1}{k^{1/2}} \left( 1 - \frac{1}{z} \right) .
\end{equation}
Furthermore, if we assume that the time of collapse occurs during the early stages of the inflationary era, i.e., $|z| \gg 1$, then 
\begin{equation}\label{INN}
 I_{NN} \simeq \frac{1}{q_1 p^{1/2} |\vq_1 - \vp|^{1/2}} \sqrt{\frac{2}{\pi}} z \sin z.
\end{equation}

Using Eqs. \eqref{PStensor11} and \eqref{INN} we can give an estimate for the tensor power spectrum within the Newtonian collapse scheme
\begin{equation}
P_h(q_1)^{\textrm{Newt}} \simeq \frac{H^4}{M_P^4 q_1^8} z^2 \sin^2 z \int d^3 p \: \frac{p_1^2 p_2^2}{p |\vq_1-\vp| }.
\end{equation}
The next step is to perform the integral over $p$ in the former expression. We can change variables once more as $\kappa_1 = u q_1$ and $p = v q_1$ to obtain
\begin{eqnarray}
P_h(q_1)^{\textrm{Newt}} &\simeq& \frac{H^4}{M_P^4 q_1^8} z^2 \sin^2 z \frac{\pi q_1^5}{64} \int_0^{v_m} dv \int_{|1-v|}^{|1+v|} du \: [4v^2-(u^2-v^2-1)^2]^2.\nn
&=& \frac{H^4}{M_P^4 q_1^8} z^2 \sin^2 z \frac{\pi q_1^5}{64} \left[ \frac{1216}{525} + \frac{256}{1575} \left(-16 + 5v_m-10 v_m^3 +21v_m^5 \right) \right].
\end{eqnarray}
Returning to the original variable $v_m = p_{UV}/q_1$, where $p_{UV}$ is the UV scale cut-off, and ignoring numerical factors, the estimated amplitude for the tensor power spectrum is finally
\begin{equation}\label{PStensor12}
 P_h(q_1)^{\textrm{Newt}} \simeq \frac{H^4}{M_P^4 q_1^8} z^2 \sin^2 z p_{UV}^5.
\end{equation}
We can compare this expression with the corresponding one obtained in the CSL case, Eq. \eqref{PStensor10}. Ignoring numerical factors and using the fact that the CSL parameter must be of the form $\lambda_{q_1} = \lambda_0/q_1 $ (due to the requirement that the scalar spectrum must be scale invariant), the estimated amplitude of the tensor power spectrum, using the CSL model, is
\begin{equation}\label{PStensor13}
 P_h(q_1)^{\textrm{CSL}} \simeq \frac{H^4}{M_P^4 q_1^6} \lambda_0^2 T^4 p_{UV}^5.
\end{equation}
As we can see, both expressions share the same structure, in particular they exhibit the same dependence on the $p_{UV}$ cut-off, and a similar increase in power at large angles ( low values of $q_1$ ). Eqs. \eqref{PStensor12} and \eqref{PStensor13} are the main results of this section.

At this point it is worth discussing the significance and implications of the aforementioned cut-off at $p_{UV}$, which appears in Eqs. \eqref{PStensor12} and \eqref{PStensor13}. The matter field under consideration is a simple scalar field with a potential and is thus fully renormalizable (as long as the potential is a polynomial of degree 4 or less). In fact, when the potential, as in the present specific case, is quadratic, we have, a free theory and thus, in principle, we are even free of the need to consider renormalization at all, except of course for the fact that composite operators such as the energy-momentum tensor needs to be renormalized anyway (something that in this situation of high symmetry can be achieved essentially via a suitable normal ordering prescription). As a result of all this, the cut-off $p_{UV}$ is not related to an ordinary renormalization issue; instead, the reason for the need to impose it has to do with the fact that the scheme we have considered involves the collapsing, and thus, the excitation, in principle, of all modes of the quantum field, including arbitrarily high UV modes.

Such modes correspond to very high values of $q \equiv | \vec q |$ and thus one might think they have nothing to do with observable quantities. The problem, however, is that, in the second order calculation, when considering the effect for a given $\vec q$, the contributions comes from two modes $\vec p$ and $ \vec q - \vec p$ (see Eq. \eqref{defsource}) and that involves modes with arbitrarily high values of $| \vec p |$ and $| \vec q - \vec p|$ (i.e. two very large vectors that add up to a small vector). That is the source of the divergence reflected in the $p_{UV}$ cut-off, as is the fact that the modification of the state of the quantum field on arbitrary high wave number modes plays a role in the relavant quantity for low wave number modes.

In order to avoid a catastrophic result we are forced to assume that, either the collapse dynamics does not affect arbitrarily high UV modes, or that something dilutes their effect. As we will discuss in more detail in the next section, we will consider the second option as a viable resultion of the issue in the present context. On the other hand, it is clear (from within the general point of view underlying this work) that modes with relatively small values of $| \vec q |$ must undergo a collapse so as to seed the scalar metric perturbations corresponding to the primordial curvature perturbations which we do observe, and those must clearly contribute to the process under consideration. We, nontheless acknowledge the fact that the issue is clearly deserving of further study, and note that it might in fact provide further clues regarding the characteristics of a viable fundamental collapse theory (i.e., it might, for instance, restrict the candidates of collapse generating operators of a general fundamental theory of spontaneous collpase), an exploration which we hope to undertake in future works.\footnote{We believe this problem is related to something similar that was uncovered, and briefly discussed, in Section XB1 of \cite{ppd}}

%%%%%%%%%%%%%%%%%%%%%%%%%%%%%%%%%%%%%%%%%%%%%%%%%%%%%%%%%%%%%
%%%%%%%%%%%%%%%%%%%%%%%%%%%%%%%%%%%%%%%%%%%%%%%%%%%%%%%%%%%%%
\section{Estimates of the B-mode polarization spectrum}
\label{predictions}
%%%%%%%%%%%%%%%%%%%%%%%%%%%%%%%%%%%%%%%%%%%%%%%%%%%%%%%%%%%%%
%%%%%%%%%%%%%%%%%%%%%%%%%%%%%%%%%%%%%%%%%%%%%%%%%%%%%%%%%%%%%

Our main results are shown in Eqs. \eqref{PStensor12} and \eqref{PStensor13}. It will be convenient to work with the dimensionless power spectrum defined as $\mP_h(q) \equiv q^3 P_h(q)$; therefore, Eqs. \eqref{PStensor12} and \eqref{PStensor13} become
\begin{equation}\label{PStensorfinal1}
 \mP_h(q_1)^{\textrm{CSL}} \simeq \frac{H^4}{M_P^4 q_1^3} \lambda_0^2 T^4 p_{UV}^5,
\end{equation}
\begin{equation}\label{PStensorfinal2}
 \mP_h(q_1)^{\textrm{Newt}} \simeq \frac{H^4}{M_P^4 q_1^5} z^2 \sin^2 z p_{UV}^5.
\end{equation}
On the other hand, in previous works, the (dimensionless) scalar power spectrum was obtained within the CSL inflationary approach \cite{ppd} and the Newtonian collapse scheme \cite{pss}. The resulting expressions are 
\begin{equation}\label{PSscalarfinal1}
 \mP_s(q_1)^{\textrm{CSL}} \simeq \frac{H^2 \lambda_0 T}{M_P^2 \epsilon},
\end{equation}
\begin{equation}\label{PSscalarfinal2}
 \mP_s(q_1)^{\textrm{Newt}} \simeq \frac{H^2}{M_P^2 \epsilon} \left( \cos z - \frac{\sin z}{z} \right)^2.
\end{equation}
Note that the scalar spectra are scale invariant. Noting that for the Newtonian scheme we assumed that the collapse occurs at the very early stages of the inflationary epoch, i.e., $|z| \gg 1$, by combining Eqs. \eqref{PSscalarfinal1}, \eqref{PSscalarfinal2}, \eqref{PStensorfinal1} and \eqref{PStensorfinal2} we obtain
\begin{equation}\label{PStensorfinal3}
 \mP_h(q_1)^{\textrm{CSL}} \simeq \epsilon^2 \frac{T^2 p_{UV}^5}{q_1^3} ( \mP_s^{\textrm{CSL}} )^2,
\end{equation}
\begin{equation}\label{PStensorfinal4}
 \mP_h(q_1)^{\textrm{Newt}} \simeq \epsilon^2 z^2 \frac{ p_{UV}^5}{q_1^5} ( \mP_s^{\textrm{Newt}} )^2,
\end{equation}
where we used the order of magnitude estimate $\sin^2 z \simeq \cos^2 z$. 

Now let us recall that the standard expression for the tensor power spectrum is usually given in terms of the so-called tensor-to-scalar ratio, $r \equiv \mP_h / \mP_s$. In the traditional inflationary paradigm, one obtains $r \simeq \epsilon$; consequently, 
\begin{equation}\label{PSstd}
\mP_h^{\textrm{std}} \simeq \epsilon \mP_s^{\textrm{std}}.
\end{equation}
Therefore, in contrast to the linear relation in both the slow-roll parameter and the scalar spectrum obtained within the standard scheme, Eq. \eqref{PSstd}, the prediction for $\mP_h$, within both collapse approaches considered is quadratic in $\epsilon$ and in $\mP_s$, Eqs. \eqref{PStensorfinal3}, \eqref{PStensorfinal4}. Another important feature of the resulting tensor power spectra of Eqs. \eqref{PStensorfinal3}, \eqref{PStensorfinal4} is that the dependence on the UV cut-off is exactly the same, i.e., they both scale as $p_{UV}^5$. All this lead us to conclude that our predictions are rather robust results of objective collapse models, as applied to the inflationary universe within a semi-classical approximation. 

%%%%%%%%%%%%%%%%%%%%%%%%%%%%%%%%%%%%%%%%%%%%%%%%%%%%%%%%%%%%%
\subsection{The magnitude of the B-mode polarization spectrum}
%%%%%%%%%%%%%%%%%%%%%%%%%%%%%%%%%%%%%%%%%%%%%%%%%%%%%%%%%%%%%

In order to estimate the magnitude of our tensor power spectra we will make the following considerations. We will evaluate them at the pivot scale $q_* = 0.05$ Mpc$^{-1}$ used by the Planck collaboration to measure the amplitude of the scalar spectrum, where it is found that $\mP_s \simeq 10^{-9}$. In addition, we observe that the obtained results depend on the physical cut-off $p_{UV}$. If we are interested in the value of the spectrum at the end of inflation, then we would assume $p_{UV}$ to be given by the last scale that exits the horizon during inflation. Nevertheless, given that we are interested in the tensor modes that might be observed in the CMB, we have to take into account the fact that primordial gravitational waves evolve through the plasma era, where they would be affected by plasma damping effects. As a result, it is reasonable to use as the effective value of the cut-off the scale of the diffusion or Silk damping, \cite{dodelson}. Assuming the two-fluid approximation of Seljak \cite{seljak}, the Silk damping scale is given by $p_{UV} = 0.078$ Mpc$^{-1}$. 

In the case of the CSL inflationary approach, we need to specify the value of $-T$, i.e., the conformal time of the beginning of inflation. Assuming 60 e-folds for the duration of inflation and an energy scale of approximately $10^{-4} M_P$ leads to $T=10^{4}$ Mpc. Inserting all the numerical values into Eq. \eqref{PStensorfinal3} we obtain 
\begin{equation}\label{PStensorfinal5}
 \mP_h(q_*)^{\textrm{CSL}} \simeq 10^{-12} \epsilon^2.
\end{equation}

In order to estimte the magnitude of the spectrum within the Newtonian collapse scheme we need to assume a particular time of collapse $\tc$ (recall that the time of collapse must be such that $\tc = z/k$). In \cite{micol2}, several values of $z$ where tested using data from the CMB. Taking into account that Eq. \eqref{PStensorfinal4} was obtained for $|z| \gg 1$, we can take $|z| = 10^{3}$, which, according to \cite{micol2}, is a value consistent with the data. Consequently, inserting all the aforementioned numerical values into Eq. \eqref{PStensorfinal4} yields
\begin{equation}\label{PStensorfinal6}
 \mP_h(q_*)^{\textrm{Newt}} \simeq 10^{-13} \epsilon^2.
\end{equation}
Clearly, these two approaches lead to rather similar, small estimates for the amplitude of the tensor power spectrum.

%%%%%%%%%%%%%%%%%%%%%%%%%%%%%%%%%%%%%%%%%%%%%%%%%%%%%%%%%%%%%
\subsection{The shape of the B-mode polarization spectrum}
%%%%%%%%%%%%%%%%%%%%%%%%%%%%%%%%%%%%%%%%%%%%%%%%%%%%%%%%%%%%%

For the sake of completeness, we also analyze the shape of our predicted B-mode polarization spectrum and compare it with the standard prediction. We begin by specifying the inflationary parameters. The predicted tensor spectra are shown in Eqs. \eqref{PStensorfinal3} and \eqref{PStensorfinal4}. Therefore, in both models, we can parametrize the spectrum as
\begin{equation}
 \mP_h(q) = A_t q^{n_t}.
\end{equation}
In the CSL model, the tensor amplitude and the tensor spectral index are given respectively by 
\begin{equation}
A_t^\CSL = T^2 p_{UV}^5 \epsilon^2 ( \mP_s^{\textrm{CSL}} )^2, \qquad n_t^\CSL = -3.
\end{equation}
On the other hand, in the Newtonian collapse scheme, the corresponding amplitude and spectral index are
\begin{equation}
A_t^\Nt = z^2 p_{UV}^5 \epsilon^2 ( \mP_s^{\textrm{Newt}} )^2, \qquad n_t^\Nt = -5.
\end{equation}
Hence, we have two different predictions for the quantities $A_t$ and $n_t$. Moreover, it is worthwhile to recall the standard prediction for the same observables
\begin{equation}
A_t^{\textrm{std}}= \epsilon \mP_s^{\textrm{std}}, \qquad n_t^{\textrm{std}} = -2\epsilon.
\end{equation}

The B-mode polarization spectrum is related to the primordial tensor power spectrum in a very similar manner as in the case of the temperature angular spectrum, i.e.,
\begin{equation}
 C_l^\BB = (4\pi)^2 \int \frac{dq}{q} \mP_h (q) \Delta_{\BB l}^2 (q),
\end{equation}
where $\Delta_{\BB l} (q)$ is the transfer function for the B-modes. These functions are obtained by integrating the resulting Boltzmann equations associated to the polarization of the CMB. 
\begin{figure}[t!]
\begin{center}
\includegraphics[scale=1.25]{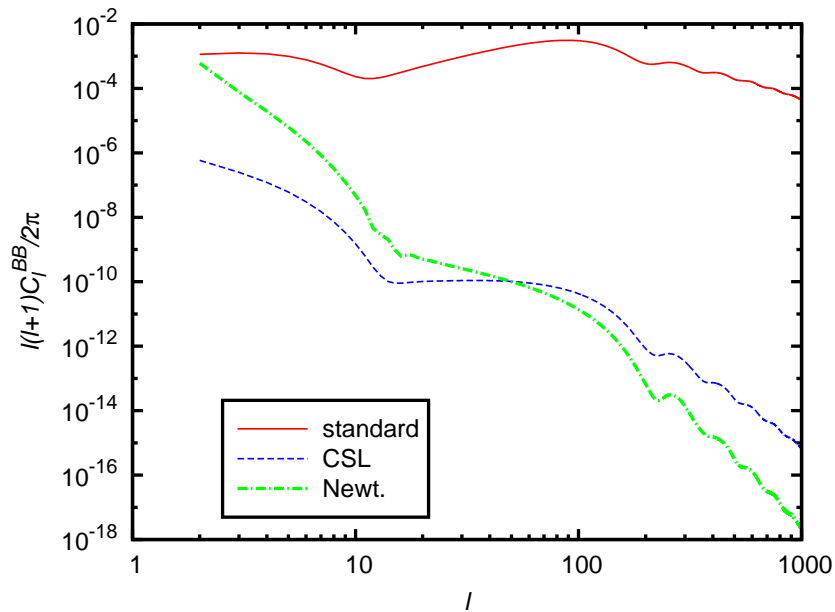}
\end{center}
\caption{The predicted B-mode polarization spectrum in three cases: the standard approach, the CSL inflationary collapse model and the Newtonian collapse scheme. In all cases, we set $\epsilon \simeq 10^{-2}$. In the standard approach that value for the slow-roll parameter is consistent with a tensor to scalar ratio $r = 0.12$.}
\label{plotClBB}
\end{figure}
In order to perform our analysis, we use a modification of the public available CAMB code, \cite{camb}. The cosmological parameters of our fiducial flat $\Lambda$CDM model considered are: baryon density in units of the critical density $\Omega_{\textrm{b}} h^2 = 0.02225$, dark matter density in units of the critical density $\Omega_{\textrm{cdm}} h^2 = 0.1198$, Hubble constant $H_0 = 67.27$ km s$^{-1}$ Mpc$^{-1}$ and reionization optical depth $\tau = 0.079$. Those are the best-fit values presented by the latest data release by the Planck mission \cite{planck2015}. 

Concerning the inflationary parameters, we choose the amplitude of the scalar spectrum to be $\mP_s \simeq 10^{-9}$ at the pivot scale $q_P = 0.05$ Mpc$^{-1}$. In addition, we set $\epsilon = 10^{-2}$. We choose this value because, according to the standard prediction, this is the order of magnitude for $\epsilon$ that would result in $r=0.12$, which saturates the highest bound set by the Planck satellite, \cite{planck2015inflation}. In this way, we can compare the predictions of our model with the standard treatment using a value that, in principle, would maximize the signal associated to the B-modes consistent with the data. On the other hand, regarding the collapse parameters, we choose the same values as before, namely, $T=10^4$ Mpc, $|z| = 10^3$ and $p_{UV} = 0.078$ Mpc$^{-1}$. As a result, we have the following amplitudes and spectral indexes
\begin{equation}
 A_t^\CSL \simeq 10^{-20}, \qquad n_t^\CSL = -3,
\end{equation}
\begin{equation}
A_t^\Nt \simeq 10^{-22} \qquad n_t^\Nt = -5.
\end{equation}
Meanwhile, for the standard prediction we have $A_t^{\textrm{std}} \simeq 10^{-11}$ and $n_t^{\textrm{std}} = -0.02$. 

In Fig. \ref{plotClBB} we present three plots of the predicted B-mode polarization spectrum. The first plot corresponds to the standard approach. We observe that this plot exhibits an amplitude and shape of the B-mode spectrum that is the same as the one shown in Fig. 14 of \cite{bicep2}, where a supposed detection of primordial B-modes was announced (and latter withdrawn \cite{dust1,dust2,dust3,dust4}). The rest of the plot presents the predicted $C_l^\BB$ from the CSL inflationary collapse model and the Newtonian collapse scheme. It is clear that our predicted amplitude is very small compared with the standard one, which was apparently in the detection range of the BICEP2 experiment, \cite{bicep2}. Note that on the multipole range $10 \lesssim l \lesssim 100$, where the standard prediction has its strongest signal, both collapse models indicate strong suppression of the B-mode spectrum with respect to the standard prediction. It is important to stress that, in generating the plots of Fig. \ref{plotClBB}, we have used the value $\epsilon = 10^{-2}$ that, according to standard predictions, corresponds to the maximum value consistent with the data. It is clear that, when the point of view we are advocating is adopted, a much weaker constraint on $\epsilon $ is implied by the data. This smaller value of $\epsilon$ will suppress even more our predicted amplitude since, in our approach, the amplitude goes as $\epsilon^2$.

We conclude that the generic prediction regarding primordial gravity waves within a self-induced collapse proposal in a semi-classical setting has an amplitude essentially undetectable by current experiments. It is worth noting, however, that our analysis indicates a rather interesting option regarding the search from primordial gravity waves. Specifically, our results imply that the search for B-modes has a higher chance of success at the largest possible angular scales (lowest values of $q'$s) shown to the left of Fig 1. In fact, if the estimate of the UV cut-off $ P_{UV}$ is increased by even a factor of 3, and we focus on the Newtonian model, the estimates suggest a possibility of seeing B-modes at, say, $ l \sim 10$. Of course, we have nothing to say regarding the technical difficulties that such a search might involve.

%%%%%%%%%%%%%%%%%%%%%%%%%%%%%%%%%%%%%%%%%%%%%%%%%%%%%%%%%%%%%
%%%%%%%%%%%%%%%%%%%%%%%%%%%%%%%%%%%%%%%%%%%%%%%%%%%%%%%%%%%%%
\section{Conclusions}
\label{concl}

In this work we have obtained the primordial tensor power spectrum within a semi-classical gravity context involving self-induced collapses of the inflaton wave function. We have used two different models for the self-induced collapse: i) an adaptation of a CSL-type model to the inflationary setting and ii) a toy model based on a single, spontaneous collapse of the initial quantum state. Given that, at first order in the perturbations, there are no sources for the tensor modes, the second order had to be considered. In both collapse models, the resulting prediction for the amplitude of the primordial tensor modes is given by $\mP_h \sim \epsilon^2 (\mP_s)^2$, i.e., it is quadratic in the scalar spectrum amplitude and in the slow-roll parameter $\epsilon$. Consequently, as shown in Fig. \ref{plotClBB}, the predicted amplitude of the B-mode polarization spectrum is several orders of magnitude smaller than the standard prediction, considering reasonable values for the cosmological and inflationary parameters. Moreover, our prediction is consistent with the latest bounds from the BICEP/Planck collaborations \cite{dust4,dust5}. Our analysis also suggests a search for B-modes at the largest possible angular scales a potentially rewarding option (if the technical difficulties of such search can be overcome).

We conclude that the current failure to detect B-modes in the CMB does not rule-out any inflationary models. It is fair to say that our specific predictions depend on some particularities of the inflationary model and the scale used as the UV cut-off. However, as seen from the fact that the two very distinct collapse models where considered, the results are rather generic. 

We acknowledge that the strong suppression of the tensor modes we find here is tied to the reliance on semi-classical treatment and not simply to the collapse hypothesis. In fact, one might in principle consider the issue by quantizing both, the perturbations of the scalar field and those of the metric, and subjecting both to a spontaneous collapse dynamics. The results in such case will naturally depend on what one assumes regarding the relationship between the part of the collapse dynamics that controls matter and metric perturbations. Generically, though, one should not expect the collapse mechanism to have the same effect on geometric and matter degrees of freedom. In fact, it is even reasonable to consider spontaneous collapse scenarios in which the geometric variables do not undergo spontaneous collapse by themselves.

A final important lesson to be drawn from this analysis is that it displays how, at least in applications to cosmology, quantum interpretational considerations can lead to dramatic modifications regarding observational issues. It thus contributes to oppose an attitude that regards such questions as of mere philosophical interest and dismisses their relevance regarding physical predictions.

%%%%%%%%%%%%%%%%%%%%%%%%%%%%%%%%%%%%%%%%%%%%%%%%%%%%%%%%%%%%%
%%%%%%%%%%%%%%%%%%%%%%%%%%%%%%%%%%%%%%%%%%%%%%%%%%%%%%%%%%%%%
\acknowledgements
%%%%%%%%%%%%%%%%%%%%%%%%%%%%%%%%%%%%%%%%%%%%%%%%%%%%%%%%%%%%%
%%%%%%%%%%%%%%%%%%%%%%%%%%%%%%%%%%%%%%%%%%%%%%%%%%%%%%%%%%%%%
G.L. acknowledges financial support from CONICET, Argentina. 
DS acknowledges partial financial support from the grants CONACYT No. 101712, and PAPIIT- UNAM No. IG100316 México, as well as sabbatical fellowships from PASPA-DGAPA-UNAM-México, and from Fulbright-Garcia Robles-COMEXUS. A. M. is supported by the DGAPA postdoctoral fellowship of UNAM. E. O. is supported by UNAM-PAPIIT grant IG100316.

%%%%%%%%%%%%%%%%%%%%%%%%%%%%%%%%%%%%%%%%%%%%%%%%%%%%%%%%%%%%%
%%%%%%%%%%%%%%%%%%%%%%%%%%%%%%%%%%%%%%%%%%%%%%%%%%%%%%%%%%%%%
\appendix 
%%%%%%%%%%%%%%%%%%%%%%%%%%%%%%%%%%%%%%%%%%%%%%%%%%%%%%%%%%%%%
%%%%%%%%%%%%%%%%%%%%%%%%%%%%%%%%%%%%%%%%%%%%%%%%%%%%%%%%%%%%%

%%%%%%%%%%%%%%%%%%%%%%%%%%%%%%%%%%%%%%%%%%%%%%%%%%%%%%%%%%%%%
%%%%%%%%%%%%%%%%%%%%%%%%%%%%%%%%%%%%%%%%%%%%%%%%%%%%%%%%%%%%%
\section{Calculations of section \ref{secPSCSL}}
\label{app1}
%%%%%%%%%%%%%%%%%%%%%%%%%%%%%%%%%%%%%%%%%%%%%%%%%%%%%%%%%%%%%
%%%%%%%%%%%%%%%%%%%%%%%%%%%%%%%%%%%%%%%%%%%%%%%%%%%%%%%%%%%%%

In this Appendix we provide a sketch of the computational steps that led to the results mentioned in section \ref{secPSCSL}. As we observe in Eq. \eqref{productosS3}, we are interested in the ensemble average $\avg{\Omega(\vkappa_1,\eta') \Omega(\vp,\eta'') \Omega(\vkappa_2,\eta''')^* \Omega (\vp',\eta'''')^* }.$ We will simplify the notation by denoting $\Omega_1 \equiv \Omega(\vkappa_1,\eta')$ , $\Omega_2 \equiv \Omega(\vp,\eta'')$, $\Omega_3 \equiv \Omega(\vkappa_2,\eta''')$, $\Omega_4 \equiv \Omega(\vp',\eta'''')$; thus 
\begin{equation}\label{defavgOmega}
 \avg{ 
 \Omega(\vkappa_1,\eta') \Omega(\vp,\eta'') \Omega(\vkappa_2,\eta''')^* \Omega 
 (\vp',\eta'''')^* } \equiv \avg{\Omega_1 \Omega_2 \Omega_3^* \Omega_4^*} .
\end{equation} 
In order to proceed, we use ``Wick's theorem'' \cite{wein}; the resulting average is therefore
\begin{equation}
 \avg{\Omega_1 \Omega_2 \Omega_3^* \Omega_4^*} = \avg{\Omega_1 \Omega_2} \cdot 
\avg{\Omega_3^* \Omega_4^*} + \avg{\Omega_1 \Omega_3^*} \cdot 
\avg{\Omega_2 \Omega_4^*} + \avg{\Omega_1 \Omega_4^*} \cdot 
\avg{\Omega_2 \Omega_3^*}.
\end{equation} 

Separating $\Omega(\nk,\eta) = w_R(\nk,\eta) + i w_I (\nk,\eta)$ and keeping in mind that $\avg{w_R(\nk, \eta) 
w_I(\nk', \eta')}=0$, i.e. the real and imaginary parts of $\Omega$ are uncorrelated, we have the following results: $\avg{ \Omega(\nk,\eta) \Omega(\nk',\eta')^* } = 2 \avg{w_\beta(\nk, \eta) w_\beta(\nk', \eta')}$; $\beta=R,I$ and $\avg{ \Omega(\nk,\eta) \Omega(\nk',\eta') } 0 = \avg{ \Omega(\nk,\eta)^* \Omega(\nk',\eta')^* }.$ Those results imply that
\begin{equation}\label{Omegasnumeros}
 \avg{\Omega_1 \Omega_2 \Omega_3^* \Omega_4^*} = 4 [\avg{w_{\beta 1} w_{\beta 3}} 
\cdot 
 \avg{w_{\beta 2} w_{\beta 4}} + \avg{w_{\beta 1} w_{\beta 4}} \cdot 
 \avg{w_{\beta 2} w_{\beta 3}}].
\end{equation} 
Moreover, the resulting expression for the ensemble averages $\avg{w_{\beta i} w_{\beta j}} $ (with $i,j = 1,2,3,4$), which was found in \cite{ppd}, is
\begin{equation}\label{omegasabhi}
 \avg{w_\beta(\nk,\eta) w_{\beta'} (\nk',\eta')} = \frac{1}{2} 
\delta_{\beta,\beta'} 
 [\delta_{\nk,\nk'} + \delta_{\nk,-\nk'} ] K(k,k';\eta,\eta'),
\end{equation} 
 where 
 \begin{equation}\label{K}
 K(k,k';\eta,\eta') \equiv \lambda_k [\delta(\eta-\eta') + M_{k} (\eta-\eta') 
\Theta(\eta-\eta') + M_{k'} (\eta'-\eta) \Theta(\eta'-\eta)+ D(k,k';\eta,\eta')];
\end{equation} 
$\Theta(x)$ is the Heaviside function, and 
\begin{equation}\label{Sk}
 S_k \equiv \frac{\lambda_k k \sqrt{2}}{\sqrt{1+\sqrt{1+4\lambda_k^2}}}
\end{equation}
\begin{equation}\label{M}
 M_{k} (\eta-\eta') \equiv \frac{2 S_k}{k} \bigg\{ S_k \sin[k(\eta-\eta')] + k 
\cos[k(\eta-\eta')] \bigg\},
\end{equation} 
\begin{equation}\label{Dk}
 D(k,k';\eta,\eta') \equiv \int_{-T}^{\eta} d\tilde \eta_1 \int_{-T}^{\eta'} d\tilde 
 \eta_2 M_k (\eta-\tilde \eta_1) M_{k'} (\eta'-\tilde \eta_2) \delta(\tilde \eta_1 - 
\tilde \eta_2). 
\end{equation} 
In fact, the quantity $-S_k$ corresponds to the imaginary part of $\alpha_k \equiv k \sqrt{1-2i\lambda_k}$, which appears in the expectation value of the field [see Eq. \eqref{expecy}]. Now, plugging in \eqref{omegasabhi} into \eqref{Omegasnumeros} and returning all the explicit indexes, we have
\begin{eqnarray}\label{avgOmega2}
 & & \avg{ \Omega(\vkappa_1,\eta') \Omega(\vp,\eta'') \Omega(\vkappa_2,\eta''')^* \Omega 
 (\vp',\eta'''')^* } \nn
& =& 
K(\kappa_1,\kappa_2;\eta',\eta''')K(p,p';\eta'', \eta'''') [\delta_{\vkappa_1,\vkappa_2} 
\delta_{\vp,\vp'} + \delta_{\vkappa_1,\vkappa_2} 
\delta_{\vp,-\vp'}+ \delta_{\vkappa_1,-\vkappa_2} 
\delta_{\vp,\vp'} + \delta_{\vkappa_1,-\vkappa_2} 
\delta_{\vp,-\vp'} ] \nn
&+& K(\kappa_1,p';\eta',\eta'''')K(p,\kappa_2;\eta'',\eta''')[\delta_{\vkappa_1,\vp'} 
\delta_{\vp,\vkappa_2} + \delta_{\vkappa_1,\vp'} 
\delta_{\vp,-\vkappa_2} + \delta_{\vkappa_1,-\vp'} 
\delta_{\vp,\vkappa_2} + \delta_{\vkappa_1,-\vp'} 
\delta_{\vp,-\vkappa_2} ] .
\end{eqnarray} 

The next task is to substitute Eq. \eqref{avgOmega2} into Eq. \eqref{productosS3} and perform the sum over $\vp'$ and make use of the $\delta$'s involving the modes $\vp'$. After summing over $\vp'$ in Eq. \eqref{productosS3}, the various terms involving the 
$\delta$'s function, will reduce to four main terms which will involve: $\delta_{\vq_1,\vq_2}$, $\delta_{\vp,(\vq_1+\vq_2)/2}$, $\delta_{\vq_1,-\vq_2}$, $\delta_{\vp,(\vq_1-\vq_2)/2}$. Next, by performing the sum over $\vp$, the terms involving $\delta_{\vp,(\vq_1+\vq_2)/2}$, $\delta_{\vp,(\vq_1-\vq_2)/2}$ will vanish exactly because the components $(1)$ and $(2)$ of the two vectors $\vq_1$ $\vq_2$ are zero (one has to actually compute the terms, as it is not entirely obvious). The only surviving terms are those which involve $\delta_{\vq_1,\vq_2}$, $\delta_{\vq_1,-\vq_2}$. After a long calculation, Eq. \eqref{productosS3} becomes
\begin{eqnarray}
 & & \avg{S(\nq_1,\eta_1 S(\nq_2,\eta_2)^* }= \frac{H^4}{2^7 M_P^4} \sum_{\vp} 
 \int_{-T}^{\eta_1} d\eta' \int_{-T}^{\eta_1} d\eta'' \int_{-T}^{\eta_2} d\eta''' 
 \int_{-T}^{\eta_2} d\eta'''' \nn
 &\times& \frac{p_1^2 p^2_2 (\delta_{\vq_1,\vq_2} + \delta_{\vq_1,-\vq_2})}{p^4 
\kappa_1^4 (\textrm{Re}[A_p])^2 (\textrm{Re}[A_{\kappa_1}])^2} 
K(p,p;\eta'',\eta'''')K(\kappa_1,\kappa_1;\eta',\eta''') \nn
&\times& \eta_1^2 \eta_2^2 
F_{\kappa_1} (\eta_1,\eta') F_{p} (\eta_1,\eta'') 
F_{\kappa_1}^* (\eta_2,\eta''') F_{p}^* (\eta_2,\eta'''') . \nn 
\end{eqnarray}
That last expression is equivalent to Eq. \eqref{productosS4}.

In the remaining part of this appendix, we show how to explicitly obtain the integral defined in Eq. \eqref{defIXX}, 
\begin{equation}\label{defIFF}
 I_{FF} (k;\eta_1,\eta_2) \equiv \int_{-T}^{\eta_1} d\eta' \int_{-T}^{\eta_2} d\eta'' 
 K(k,k;\eta',\eta'') F_k (\eta_1,\eta') F_k^* (\eta_2,\eta'').
\end{equation} 
Actually, the direct calculation of $I_{FF}$ is not a simple task due, in particular, to the fact that the function $K(k,k;\eta',\eta'')$ [see definition \eqref{Dk}], is in itself a nontrivial double integral. We will circumvent the direct calculation and obtain the exact value of $I_{FF}$ by following an alternative path. 

We start by recalling the expression for the expectation value $\expec{\hat y (\nk,\eta)}$, (Eq. \eqref{expecy}) is:
\begin{eqnarray}\label{expecpi2}
 \expec{\hat y (\nk, \eta)} &=& \frac{L^{3/2}}{2^{3/2} k^2 2 \textrm{Re}[A_k]} 
 \int_{-T}^{\eta} d\eta'\: \Omega(\vk,\eta') F_k (\eta,\eta') \nn
 &=& \frac{L^{3/2}}{2^{3/2} k^2 2 \textrm{Re}[A_k]} 
 \int_{-T}^{\eta} d\eta'\: [w_S (\vk,\eta') + i w_A(\vk,\eta') ] 2 i e^{S_k(\eta'-\eta)} \bigg[ \bigg(S_k - \frac{1}{\eta} \bigg) \cos[R_k(\eta-\eta')] - R_k \sin[R_k(\eta-\eta')] \bigg] , \nn
\end{eqnarray}
where in the last line we have used the explicit expression for $F_k (\eta,\eta')$ and separated $\alpha_k \equiv k \sqrt{1-2i\lambda_k}$ into its real part $R_k$ and its imaginary part $-S_k$. Next, the noise functions $w_{R,I} (\vk,\eta)$ can be expressed in terms of new noise functions $v_{R,I} (\vk,\eta)$.
\begin{equation}
 v_{R,I} (\vk,\eta) \equiv w_{R,I} (\vk,\eta) - 2 \lambda_k \expec{\hat \pi^{\RI} (\vk,\eta)}.
\end{equation}
The advantage of the noise $v_{R,I} $ over the $w_{R,I} $ is that its variance is much easier 
to handle. In particular, we have
\begin{equation}\label{defw}
 w_\beta (\vk,\eta) = v_{\beta} (\vk,\eta) + \int_{-T}^\eta d\tilde \eta \: 
M_k(\eta',\tilde \eta) v_{\beta} (\vk,\tilde \eta) 
\end{equation} 
with 
\begin{equation}\label{avgv}
 \avg{v_\beta (\vk,\eta_1) v_{\beta'} (\vq,\eta_2) }= \frac{\lambda_k}{2}[ 
 \delta_{\vk,\vq} + \delta_{\vk,-\vq}] \delta_{\beta,\beta'} \delta (\eta_1- \eta_2) .
 \end{equation} 
Substituting \eqref{defw} into \eqref{expecpi2}, and performing one of the integrals (but exchanging the order of integration, which in turn changes the limits of integration, i.e., $
 \int_{-T}^\eta d\eta ' \int_{-T}^{\eta'} d \tilde \eta = \int_{-T}^\eta d\tilde \eta 
\int_{\tilde \eta}^\eta d \eta'$ ), yields
\begin{equation}\label{expecpifinal}
 \expec{\hat y (\nk, \eta)} = \frac{-L^{3/2} }{2^{3/2} k^2 \eta
 \textrm{Re}[A_k] } \int_{-T}^\eta d\eta' N_k (\eta,\eta')[v_R(\vk,\eta') 
+ i v_I(\vk, \eta')],
\end{equation} 
where
\begin{equation}\label{Nk}
N_k (\eta,\eta') \equiv (S_k \eta-1) \cos[k(\eta-\eta')] - \bigg(\frac{S_k}{k} + k\eta \bigg) \sin [k(\eta-\eta')] .
\end{equation}
With \eqref{expecpifinal} at hand, we can now proceed to calculate
\begin{eqnarray}
 \avg{ \expec{\hat y (\nk, \eta_1)} \expec{\hat y (\vq, \eta_2)}^* } &=& 
 \frac{L^3}{2^3 \textrm{Re}[A_k] \textrm{Re}[A_q] k^2 \eta_1 q^2 \eta_2 } \int_{-T}^{\eta_1} 
 d\eta' \int_{-T}^{\eta_2} d\eta'' N_k (\eta_1,\eta') N_q (\eta_2,\eta'') \nn
 &\times& \left[ \avg{ 
 v_R(\vk,\eta')v_R(\vq,\eta'')} + \avg{ v_I(\vk,\eta')v_I(\vq,\eta'') } \right] .
\end{eqnarray} 
Using \eqref{avgv} in the above expression we find 
\begin{eqnarray}\label{avgpipi}
 \avg{ \expec{\hat y (\nk, \eta_1)} \expec{\hat y (\vq, \eta_2)}^* } &=& 
 \frac{L^3 \lambda_k [\delta_{\vk,\vq}+\delta_{\vk,-\vq}]}{2^3 (\textrm{Re}[A_k])^2 k^4 \eta_1 \eta_2 } \int_{-T}^{\eta_1} d\eta' \int_{-T}^{\eta_2} d\eta'' N_k (\eta_1,\eta') N_k 
(\eta_2,\eta'') \delta (\eta'-\eta'').
\end{eqnarray} 

The double integral in that last expression can be performed using the formula 
%for integrating a double integral involving a Dirac's delta, see 
\eqref{intdobledelta}, which leads to
%This formula, and others, are deduced in the Appendix \ref{app2}. Thus, using formula \eqref{intdobledelta} we have
\begin{equation}
 \int_{-T}^{\eta_1} d\eta' \int_{-T}^{\eta_2} d\eta'' N_k 
(\eta_1,\eta') N_k (\eta_2,\eta'') \delta (\eta'-\eta'') = v_{FF} (k;T) - v_{FF} (k;-\eta_2) \Theta (\eta_1 
- \eta_2) - v_{FF} (k;-\eta_1) \Theta (\eta_2 - \eta_1) ,
\end{equation} 
where we have defined the function
\begin{equation}
 v_{FF} (k;z) \equiv \int^z d\zeta N_k (\eta_1,-\zeta) N_k (\eta_2,-\zeta) ,
\end{equation} 
which is explicitly given by 
\begin{eqnarray}\label{vFFdef}
 v_{FF} (k;z) &\equiv& \frac{1}{4k^3}\bigg[ k \left(2 {S_k} \left({\eta_1} {\eta_2} 
k^2-1\right)+k^2 (-({\eta_1}+{\eta_2}))+{S_k}^2 ({\eta_1}+{\eta_2})\right) \cos (k 
({\eta_1}+{\eta_2}+2 z)) \nn
&+& 2 k z \left(k^2+{S_k}^2\right) \left(\left({\eta_1} 
{\eta_2} k^2+1\right) \cos (k ({\eta_1}-{\eta_2}))+k ({\eta_1}-{\eta_2}) 
\sin (k ({\eta_1}-{\eta_2}))\right)\nn
&-&\left({\eta_1} {\eta_2} k^4+k^2 
({\eta_1} {S_k} (2-{\eta_2} {S_k})+2 {\eta_2} {S_k}-1)+{S_k}^2\right) \sin (k 
({\eta_1}+{\eta_2}+2 z))\bigg].
\end{eqnarray} 
As a consequence, \eqref{avgpipi} is given by 
\begin{eqnarray}\label{avgpipi2}
 \avg{ \expec{\hat y (\nk, \eta_1)} \expec{\hat y (\vq, \eta_2)}^* } &=& 
 \frac{L^3 \lambda_k [\delta_{\vk,\vq}+\delta_{\vk,-\vq}]}{2^3 (\textrm{Re}[A_k])^2 k^4 \eta_1 \eta_2} 
\nn
&\times& \left[v_{FF} (k;T) - v_{FF} (k;-\eta_2) \Theta (\eta_1 
- \eta_2) - v_{FF} (k;-\eta_1) \Theta (\eta_2 - \eta_1) \right].
\end{eqnarray}

On the other hand, we can once again use the original expression for the expectation value of the field \eqref{expecy} and find, 
\begin{equation}\label{avgpipi3}
 \avg{ \expec{\hat y (\nk, \eta_1)} \expec{\hat y (\vq, \eta_2)}^* }= \frac{ 
 L^{3}}{2^{3} 4 \text{Re} [A_k] \text{Re} [A_q] k^2 q^2 } \int_{-T}^{\eta_1} d\eta' 
 \int_{-T}^{\eta_2} d\eta'' \avg{ \Omega(\nk,\eta') \Omega(\vq,\eta'')^*} F_{k} 
(\eta_1,\eta') F_{q} (\eta_2,\eta'')^*. 
\end{equation}
Using that 
\begin{equation}
 \avg{ \Omega(\nk,\eta') \Omega(\vq,\eta'')^*} = [\delta_{\vk,\vq} + 
\delta_{\vk,-\vq}] K (k,q; \eta',\eta''),
\end{equation} 
 \eqref{avgpipi3} can be rewritten as
\begin{eqnarray}\label{avgpipi4}
 \avg{ \expec{\hat y (\nk, \eta_1)} \expec{\hat y (\vq, \eta_2)}^* } &=& \frac{ 
 L^{3} [\delta_{\vk,\vq} + 
\delta_{\vk,-\vq}] }{2^{5} (\text{Re} [A_k])^2 k^4 } \int_{-T}^{\eta_1} d\eta' 
 \int_{-T}^{\eta_2} d\eta'' K (k,k; \eta',\eta'') F_{k} 
(\eta_1,\eta') F_{q} (\eta_2,\eta'')^* \nn
&=& \frac{ L^{3} [\delta_{\vk,\vq} + \delta_{\vk,-\vq}] }{2^{5} (\text{Re} [A_k])^2 k^4
} I_{FF} (k;\eta_1,\eta_2)
\end{eqnarray} 
Comparing \eqref{avgpipi2} and \eqref{avgpipi4} we finally find that
\begin{equation}\label{IFF}
 I_{FF} (k;\eta_1,\eta_2) = \frac{4 \lambda_k}{\eta_1 \eta_2} \left[v_{FF} (k;T) - 
v_{FF} (k;-\eta_2) \Theta (\eta_1 - \eta_2) - v_{FF} (k;-\eta_1) \Theta (\eta_2 - \eta_1) 
\right],
\end{equation} 
with the function $v_{FF}$ defined in Eq. \eqref{vFFdef}.

%%%%%%%%%%%%%%%%%%%%%%%%%%%%%%%%%%%%%%%%%%%%%%%%%%%%%%%%%%%%%
%%%%%%%%%%%%%%%%%%%%%%%%%%%%%%%%%%%%%%%%%%%%%%%%%%%%%%%%%%%%%
\section{Estimation of the integrals of the CSL tensor power spectrum}
\label{app2}
%%%%%%%%%%%%%%%%%%%%%%%%%%%%%%%%%%%%%%%%%%%%%%%%%%%%%%%%%%%%%
%%%%%%%%%%%%%%%%%%%%%%%%%%%%%%%%%%%%%%%%%%%%%%%%%%%%%%%%%%%%%

In this appendix we provide an estimate for the integrals appearing in Eq. \eqref{PStensor6}. Let us start by defining a new function 
\begin{equation}\label{XFFdef}
 X_{FF} (u,v;x_1,x_2) \equiv \frac{x_1^2 x_2^2 uv}{q_1^4 16 \lambda_q^2} I_{FF} ( vq_1; 
x_1/q_1,x_2/q_1) I_{FF} ( uq_1; x_1/q_1,x_2/q_1);
\end{equation}
Using the result obtained in the previous appendix corresponding to $I_{FF}$ \eqref{IFF}, we find that 
\begin{eqnarray}
 X_{FF} (u,v;x_1,x_2) &=& v_{FF}(v q_1 ; T )v_{FF}(u q_1 ;T ) + [-v_{FF}(v q_1 ;T ) 
v_{FF}(u q_1 ; -x_2/q_1 ) \nn 
&+& v_{FF}( vq_1;-x_2/q_1 ) v_{FF}(uq_1 ;-x_2/q_1 ) - v_{FF}(u q_1;T ) v_{FF}(vq_1 
;-x_2/q_1 ) ]\Theta(x_1-x_2) \nn
&+& [-v_{FF} (vq_1 ;T ) v_{FF} ( uq_1 ;-x_1/q_1 ) + v_{FF} (vq_1 ;-x_1/q_1 ) v_{FF} 
( uq_1 ; -x_1/q_1 ) \nn
&-& v_{FF} (vq_1 ;-x_1/q_1 ) v_{FF} ( uq_1 ;T ) 
]\Theta(x_2-x_1).
\end{eqnarray} 

The next task is to substitute the the function $v_{FF}$ \eqref{vFFdef} in the above expression. This expression for $X_{FF} (u,v;x_1,x_2)$ is rather cumbersome, and thus we will make some approximations. First, we will only retain the leading term in 
powers of $T$. Let us recall that $T$ is in general a very large number in absolute value, since it represent the conformal time at the beginning of inflation or the conformal time where the vacuum was selected. Second, we will bound the oscillating terms in $v_{FF}$ (i.e. we use that $ |\cos[x] | \leq 1 $ and $|\sin[x]| \leq 1$ ). Thus, the approximated expression for $X_{FF}$ is given by
\begin{eqnarray}\label{XFFapp}
 X_{FF} (u,v;x_1,x_2) &\simeq& \frac{[(q_1u)^2 + S_{q_1u}^2] [(q_1v)^2 + S_{q_1v}^2]}{4 
q_1^6} \bigg[ q_1^2 T^2 x_1^2 x_2^2 \nn
&+& x_1^2 x_2^3 (2 q_1 T + x_2) \Theta (x_1-x_2) + x_2^2 x_1^3 (2 q_1 T + x_1) \Theta 
(x_2-x_1) \bigg].
\end{eqnarray} 

With Eq. \eqref{XFFapp} at hand, and with the help of Eq. \eqref{inttheta2dimfinal}, we can evaluate the following integral
\begin{eqnarray}\label{intXFF}
 & & \int_{-q_1T}^0 dx_1 \int_{-q_1T}^0 dx_2 J_{3/2}(x_1) x_1^{-1/2} J_{3/2}(x_2) 
x_2^{-1/2} X_{FF} (u,v;x_1,x_2) \simeq \frac{2 q_1^4 T^4}{ \pi } \frac{[(q_1u)^2 + 
S_{q_1u}^2] [(q_1v)^2 + S_{q_1v}^2]}{4 q_1^6} \nn
&\times& \left[ \sin^2 (q_1T) + 2 \left( - \frac{1}{3} + \frac{\cos(2 q_1 T)}{2} 
\right) + 2 \left( \frac{1}{8} - \frac{\cos(2 q_1 T)}{4} \right) \right] \nn
&=& \frac{ q_1^4 T^4}{ \pi } \frac{[(q_1u)^2 + 
S_{q_1u}^2] [(q_1v)^2 + S_{q_1v}^2]}{24 q_1^6},
\end{eqnarray}
where we have used the identity $\cos(2x) = 1 - 2 \sin^2 x$. Recalling that $R_k$ corresponds to the real part of $\alpha_k \equiv k \sqrt{1-2i\lambda_k}$, we have the relation
\begin{equation}
 2 (R_{q_1 u})^2 = {u q_1^2}\left( u + \sqrt{u^2 + 4 \lambda_{q_1}^2} \right)
\end{equation} 
Using the above result, we can rewrite the factor 
\begin{equation}\label{factor}
 \frac{[4v^2-(u^2-v^2-1)^2]^2}{uv 
 [u+(u^2+4\lambda_{q_1}^2)^{1/2}][v+(v^2+4\lambda_{q_1}^2 )^{1/2} ] } = 
\frac{[4v^2-(u^2-v^2-1)^2]^2 q_1^4}{2^2 R_{q_1 u}^2 R_{q_1 v}^2 } 
\end{equation} 

Eqs. \eqref{intXFF} and \eqref{factor} can now be used to evaluate the main integral \eqref{PStensor6}, i.e.,
 \begin{eqnarray}\label{PStensor6a}
 P_h (q_1) &=& \frac{ \pi^2 H^4 
\lambda_{q_1}^2}{ q_1 2^5 M_P^4} 
 \int_0^{\infty} dv \int_{|1-v|}^{|1+v|} du 
\frac{[4v^2-(u^2-v^2-1)^2]^2}{uv 
 [u+(u^2+4\lambda_{q_1}^2)^{1/2}][v+(v^2+4\lambda_{q_1}^2 )^{1/2} ] } \nn
&\times& \int_{-q_1T}^0 dx_1 \int_{-q_1T}^0 dx_2 \: \frac{1}{\sqrt{x_1 x_2}} 
J_{3/2} (x_1) J_{3/2} (x_2) X_{FF} (u,v;x_1,x_2)\nn
&\simeq& \frac{ \pi^2 H^4 \lambda_{q_1}^2}{ q_1 2^5 M_P^4} \frac{ q_1^4 T^4}{24 q_1^6 \pi 
 } \nn
 &\times& \int_0^{\infty} dv \int_{|1-v|}^{|1+v|} du \: q_1^4 [4v^2-(u^2-v^2-1)^2]^2 
\frac{[(q_1u)^2 + S_{q_1u}^2] [(q_1v)^2 + S_{q_1v}^2]}{2^2 R_{q_1 u}^2 R_{q_1 v}^2 }
\end{eqnarray} 
Making some simplifications in the last equation, and using the very important relation $R_k^2 - S_k^2 = k^2$, we have that
\begin{equation}\label{PStensor9}
 P_h (q_1) \simeq \frac{ \pi H^4 \lambda_{q_1}^2}{3072 q_1^3 M_P^4} q_1^4 T^4 
\int_0^{v_m} dv \int_{|1-v|}^{|1+v|} du \: [4v^2-(u^2-v^2-1)^2]^2 .
\end{equation} 

Note that we have introduced an UV cut-off $v_m$ in the integral given in \eqref{PStensor9}. With the cut-off, the integral can be done analytically and the result is 
\begin{equation}
 P_h (q_1) \simeq \frac{ \pi H^4 \lambda_{q_1}^2}{3072 q_1^3 M_P^4} q_1^4 T^4 \left[ 
\frac{1216}{325} + \frac{256}{1575} (-16 + 5 v_m -10 v_m^3 +21 v_m^5 ) \right].
\end{equation} 
Hence, the tensor power spectrum $P_h$, is proportional to the fifth power of the UV cut-off, $v_m^5$.

%%%%%%%%%%%%%%%%%%%%%%%%%%%%%%%%%%%%%%%%%%%%%%%%%%%%%%%%%%%%%
%%%%%%%%%%%%%%%%%%%%%%%%%%%%%%%%%%%%%%%%%%%%%%%%%%%%%%%%%%%%%
\section{Double integrals involving Dirac's delta and Heavisde step function}
\label{app3}
%%%%%%%%%%%%%%%%%%%%%%%%%%%%%%%%%%%%%%%%%%%%%%%%%%%%%%%%%%%%%
%%%%%%%%%%%%%%%%%%%%%%%%%%%%%%%%%%%%%%%%%%%%%%%%%%%%%%%%%%%%%

Let us begin by evaluating the following integral
\begin{equation}
 \int_{a}^b dx \: f(x) \Theta(x-x_0) = v_1 (b) \Theta(b-x_0) - v_1 (a) \Theta (a-x_0) - 
 \int_{a}^b dx\: v_1 (x) \delta(x-x_0),
\end{equation}
where we have integrated by parts using that $\frac{d}{dx} \Theta(x-x_0) = \delta (x-x_0)$ and 
\begin{equation}
 v_1 (x) \equiv \int^x dy\: f(y),
\end{equation} 
i.e., denotes the primitive (or anti-derivative) of $f(x)$. Therefore, 
\begin{equation}\label{inttheta1dim}
 \int_{a}^b dx \: f(x) \Theta(x-x_0) = v_1(b) \Theta(b-x_0) - v_1(a) \Theta (a-x_0) - 
v_1(x_0)\Theta(x_0-a)\Theta(b-x_0),
\end{equation} 
where the last term comes from performing the integral with the Dirac's delta, and the product of the $\Theta$ functions ensure that $x_0$ is within the interval $[a,b]$. Eq. \eqref{inttheta1dim} is the first main formula we use for evaluating the 
remaining integrals. 

With equation \eqref{inttheta1dim} at hand, we can evaluate the double integral involving a Dirac's delta. In particular, in our work, we are interested in evaluating integrals of the form
\begin{equation}
 \int_{-T}^{a} d x \int_{-T}^{b} dy f(x,y) \delta (x-y).
\end{equation} 
Given that the integration limits are well-defined, we can interchange them via Fubini's theorem and perform first the integral over the $x$ variable; i.e.,
\begin{eqnarray}
\int_{-T}^{a} d x \int_{-T}^{b} dy\: f(x,y) \delta (x-y) &=& \int_{-T}^{b} dy \:
\Theta(y+T) \Theta (a-y) f(y,y) \nn
&=& \int_{-T}^{b} dy \: f(y,y) \Theta(a-y) \nn
&=& \int_{-b}^T dz f(-z,-z) \Theta(z-z_0) ,
\end{eqnarray}
where in the first line, the products of the $\Theta$ functions ensure that the $y$ variable is within the interval $[-T,a]$; in the second line we have used that $\Theta(y+T) =1$ because $y \in [-T,b]$ and $-T<b$; in the third line we have made a change of variable $y=-z$ and $z_0=-a$. Using \eqref{inttheta1dim} we have
\begin{equation}
 \int_{-b}^T dz f(-z,-z) \Theta(z-z_0) = v(T)\Theta(T+a) - v(-b) \Theta(-b+a) - 
v(-a)\Theta(T+a)\Theta(-a+b),
\end{equation} 
where
\begin{equation}
 v(y) = \int^y dz f(-z,-z).
\end{equation} 
Furthermore, since $-T<a$ then $\Theta(T+a) = 1$; thus
\begin{equation}\label{intdobledelta}
 \int_{-T}^{a} d x \int_{-T}^{b} dy\: f(x,y) \delta (x-y) = v(T) - v(-b) 
\Theta(a-b) - v(-a)\Theta(b-a).
\end{equation} 
Eq. \eqref{intdobledelta} is one of the main formulas we use. 

Next, another double integral that we employ involves the Heaviside step function. In particular, we use integrals of the form
\begin{equation}
 \int_{-T}^{a} d x \int_{-T}^{b} dy f(x,y) \Theta(x-y).
\end{equation} 
In order to evaluate the previous integral, we use Fubini's theorem and perform first the integral over the $x$ variable and then use \eqref{inttheta1dim}, i.e.,
\begin{eqnarray}\label{inttheta2dim}
 \int_{-T}^{b} d y \int_{-T}^{a} dx f(x,y) \Theta(x-y) &=& \int_{-T}^{b} d y \bigg[ 
 v(a,y) \Theta(a-y) - v(-T,y) \Theta(-T-y) - v(y,y) \Theta(y+T) \Theta(a-y) \bigg] \nn
 &=& \int_{-T}^{b} d y \bigg[ 
 v(a,y) - v(y,y) \bigg] \Theta(a-y),
\end{eqnarray}
where in the last line we used that $T+y >0$, which implies $\Theta(-T-y) = 0$ and $ \Theta(T+y) = 1$; we have also defined
\begin{equation}
 v(x,y) = \int^x d\zeta\: f(\zeta,y).
\end{equation} 
Next, we perform a change of variable $y=-s$ in the last integral of \eqref{inttheta2dim}, thus 
\begin{equation}
 \int_{-T}^{b} d y \bigg[ v(a,y) - v(y,y) \bigg] \Theta(a-y) = \int_{-b}^{T} d s 
\bigg[ v(a,-s) - v(-s,-s) \bigg] \Theta(s-s_0) ,
\end{equation} 
where $s_0 = -a$. Once again, using \eqref{inttheta1dim} we find,
\begin{equation}
 \int_{-b}^{T} d s 
\bigg[ v(a,-s) - v(-s,-s) \bigg] \Theta(s-s_0) = w(a,T)\Theta(T+a)-w(a,-b)\Theta(-b+a) 
- w(a,-a) \Theta(-a+b) \Theta(T+a),
\end{equation} 
where 
\begin{equation}
 w(a,z) = \int^z d\zeta \: [v(a,-\zeta)-v(-\zeta,-\zeta)]
\end{equation} 
and, since $T+a >0$, we arrive at the final expression 
\begin{equation}\label{inttheta2dimfinal}
 \int_{-T}^{b} d y \int_{-T}^{a} dx f(x,y) \Theta(x-y) = 
w(a,T)-w(a,-b)\Theta(a-b) - w(a,-a) \Theta(b-a) 
\end{equation} 
Eq. \eqref{inttheta2dimfinal} is the final main formula we use in our work.

%%%%%%%%%%%%%%%%%%%%%%%%%%%%%%%%%%%%%%%%%%%%%%%%%%%%%%%%%%%%%
%%%%%%%%%%%%%%%%%%%%%%%%%%%%%%%%%%%%%%%%%%%%%%%%%%%%%%%%%%%%%
\section{Definition of the projection tensor $\mP_{ij}^{~~lm}$}
\label{AppPtensor}
%%%%%%%%%%%%%%%%%%%%%%%%%%%%%%%%%%%%%%%%%%%%%%%%%%%%%%%%%%%%%
%%%%%%%%%%%%%%%%%%%%%%%%%%%%%%%%%%%%%%%%%%%%%%%%%%%%%%%%%%%%%

In this section we define the projection tensor $\mP_{ij}^{~~lm}$. As we have mentioned, the tensor $\mP_{ij}^{~~lm}$ serves to extract the transverse and traceless (TT) part of any tensor. We begin by introducing the basis in which any TT tensor can be decomposed. That is, if $h_{ij}$ is a TT tensor, then its Fourier transform is
\begin{equation}
 h_{ij} (\x,\eta) = \int \frac{d^3 \vk}{(2\pi)^{3/2}} e^{i \vk \cdot \x} \left[ h_{\vk} (\eta) e_{ij} (\vk) + \tilde h_{\vk} (\eta) \tilde e_{ij} (\vk) \right],
\end{equation} 
where we defined two time-independent polarization tensors $e_{ij}$ and $\tilde e_{ij}$. The polarization tensors may be expressed in terms of orthonormal basis vectors $e_i, \tilde e_j$ and $\vk$. Explicitly
\begin{equation}
 e_{ij} (\vk) = \frac{1}{\sqrt{2}} \left[ e_i (\vk) e_j (\vk) - \tilde e_i (\vk) \tilde e_j (\vk) \right],
\end{equation} 
\begin{equation}
\tilde e_{ij} (\vk) = \frac{1}{\sqrt{2}} \left[ e_i (\vk) \tilde e_j (\vk) + \tilde e_i (\vk) e_j (\vk) \right].
\end{equation}
In terms of these polarization tensors, the action of the projection tensor $\mP_{ij}^{~~lm}$ on any tensor $S_{lm}$ is defined as
\begin{equation}
 \mP_{ij}^{~~lm} S_{lm} \equiv \int \frac{d^3 \vk}{(2\pi)^{3/2}} e^{i \vk \cdot \x} \left[ e_{ij} (\vk) e^{lm} (\vk) + \tilde e_{ij} (\vk) \tilde e^{lm} (\vk) \right] S_{lm} (\vk),
\end{equation} 
where $S_{lm}(\vk)$ is the Fourier transform of the tensor $S_{lm} (\x,\eta)$, i.e.,
\begin{equation}
 S_{lm} (\vk,\eta) = \int \frac{d^3 \x}{(2\pi)^{3/2}} e^{-i \vk \cdot \x} S_{lm} (\x,\eta).
\end{equation} 

%%%%%%%%%%%%%%%%%%%%%%%%%%%%%%%%%%%%%%%%%%%%%%%%%%%%%%%%%%%%%
%%%%%%%%%%%%%%%%%%%%%%%%%%%%%%%%%%%%%%%%%%%%%%%%%%%%%%%%%%%%%
\bibliography{bibliografia0}
\bibliographystyle{apsrev} 
%%%%%%%%%%%%%%%%%%%%%%%%%%%%%%%%%%%%%%%%%%%%%%%%%%%%%%%%%%%%%
%%%%%%%%%%%%%%%%%%%%%%%%%%%%%%%%%%%%%%%%%%%%%%%%%%%%%%%%%%%%%
\end{document}